\documentclass[twocolumn,pre,superscriptaddress,showpacs]{revtex4-1}
\usepackage{graphicx,color}
\usepackage{dcolumn}
\usepackage{amsmath,amssymb}
\usepackage{longtable}
\usepackage{mathtools}
\usepackage{cases}
\usepackage{booktabs}
\usepackage[capitalize]{cleveref}

\newcommand{\CHK}[1]{\textcolor{black}{#1}}

\newcommand{\exFig}[1]{\CHK{Fig.~\!{#1}}}
\newcommand{\exEq}[1]{\CHK{Eq.~\!{#1}}}
\newcommand{\exEqs}[1]{\CHK{Eqs.~\!{#1}}}
\newcommand{\exSec}[1]{\CHK{Sec.~\!{#1}}}
\newcommand{\exchap}[1]{\CHK{Chap.~\!{#1}}}

\newcommand{\myRef}[1]{Ref.~\!\cite{#1}}
\newcommand{\myRefs}[1]{Refs.~\!\cite{#1}}
\newcommand{\myEq}[1]{Eq.~\!(\ref{#1})}
\newcommand{\myEqs}[1]{Eqs.~\!(\ref{#1})}
\newcommand{\myEquation}[1]{Equation~\!(\ref{#1})}
\newcommand{\mySec}[1]{Sec.~\!\ref{#1}}
\newcommand{\mySecs}[1]{Secs.~\!\ref{#1}}
\newcommand{\myFig}[1]{Fig.~\!\ref{#1}}
\newcommand{\myFigs}[1]{Figs.~\!\ref{#1}}
\newcommand{\myTable}[1]{Table~\!\ref{#1}}
\newcommand{\myAppendix}[1]{Appendix~\!\ref{#1}}
\newcommand{\myav}[1]{\left\langle#1\right\rangle}

\def\myIm{{\rm i}}
\def\myTr{{\rm Tr}}
\def\mykB{k_{\rm B}}
\def\myTc{T_{\rm C}}
\def\mysn{{\rm sn}}
\def\mycn{{\rm cn}}
\def\mydn{{\rm dn}}
\def\myann{{\cal D}}

\newcommand\FIGone{%
 \begin{figure}[tb]
  \begin{center}
   \includegraphics[width=0.9\linewidth]{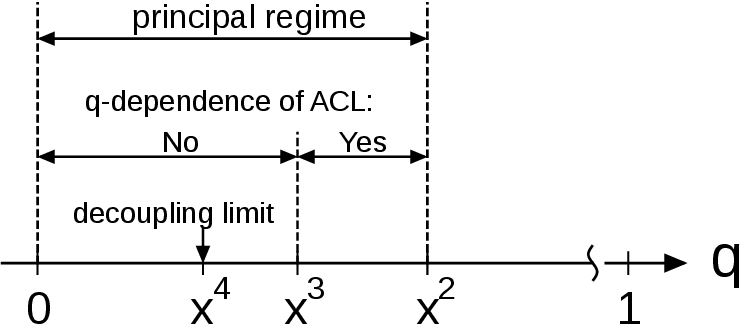}
  \end{center}
  \caption{%
  A schematic diagram of the principal regime in the eight-vertex
  model.
  For a given $x$ $(0<x<1)$,
  there exist two cases with respect to another parameter $q$:
  The ACL depends on $q$ for $x^3<q<x^2$ and does not for
  $0<q<x^3$.
  The latter region includes the decoupling limit $q=x^4$.
  }
  \label{fig:1}
 \end{figure}
 }
 
 \newcommand\FIGtwo{%
 \begin{figure}[tb]
  \begin{center}
   \includegraphics[width=\linewidth]{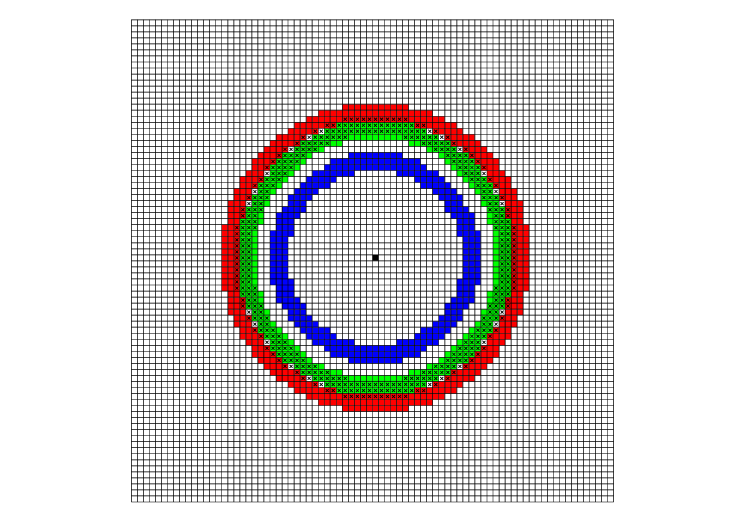}
  \end{center}
   \caption{%
  Annular regions. 
  The origin $(0,0)$ is denoted by the black cell.
  The annular region ${\myann}(10^{-3},3\times10^{-4})$
  is employed for the fitting of correlation function data of
  $Q=2$ at $t=0.24$ (308 blue cells).
  The annulus for $Q=3$ at $t=0.15$
  and
  that        for $Q=4$ at $t=0.1$
  are indicated by
  green cells
  and
  red cells,
  respectively.
  The annulus for $Q=1$ at $t=0.50$
  are
  given by
  crosses
  overwritten on the cells.
  }
  \label{fig:2}
 \end{figure}
 }
 
 \newcommand\FIGthree{%
 \begin{figure}[t]
  \begin{center}
   \includegraphics[width=0.475\linewidth]{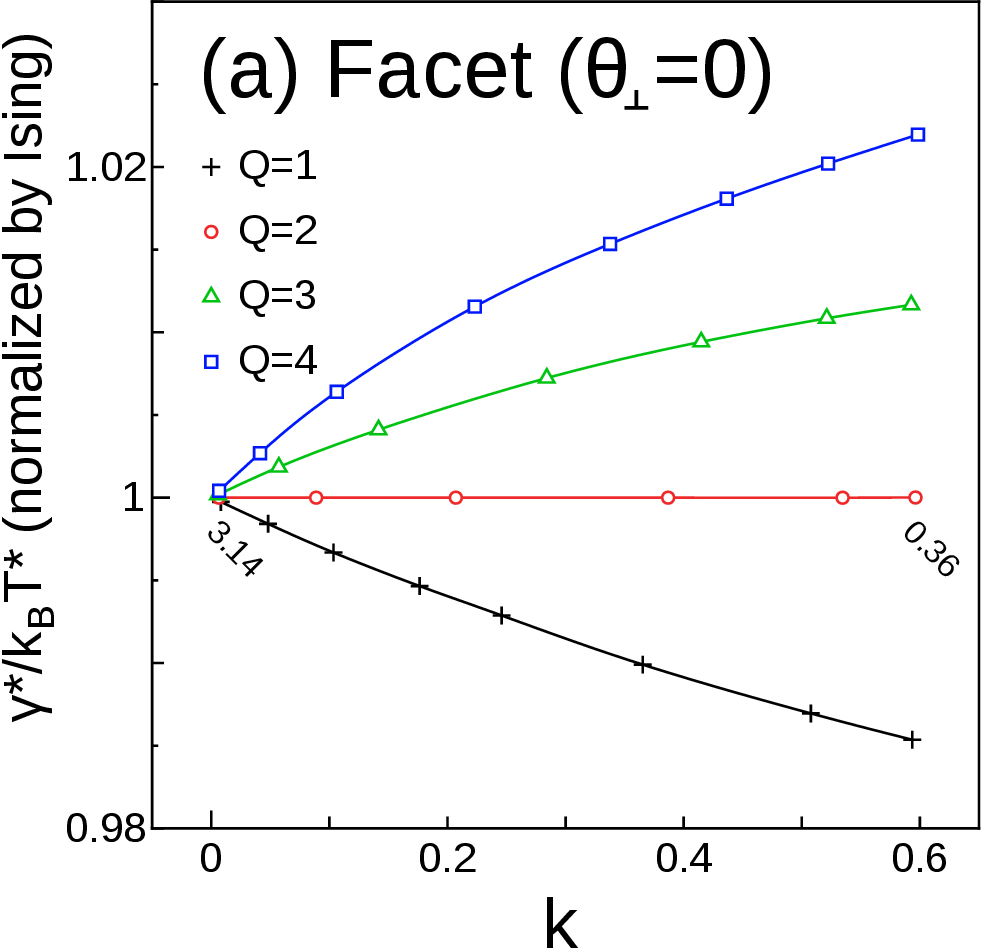}\hspace{2mm}
   \includegraphics[width=0.475\linewidth]{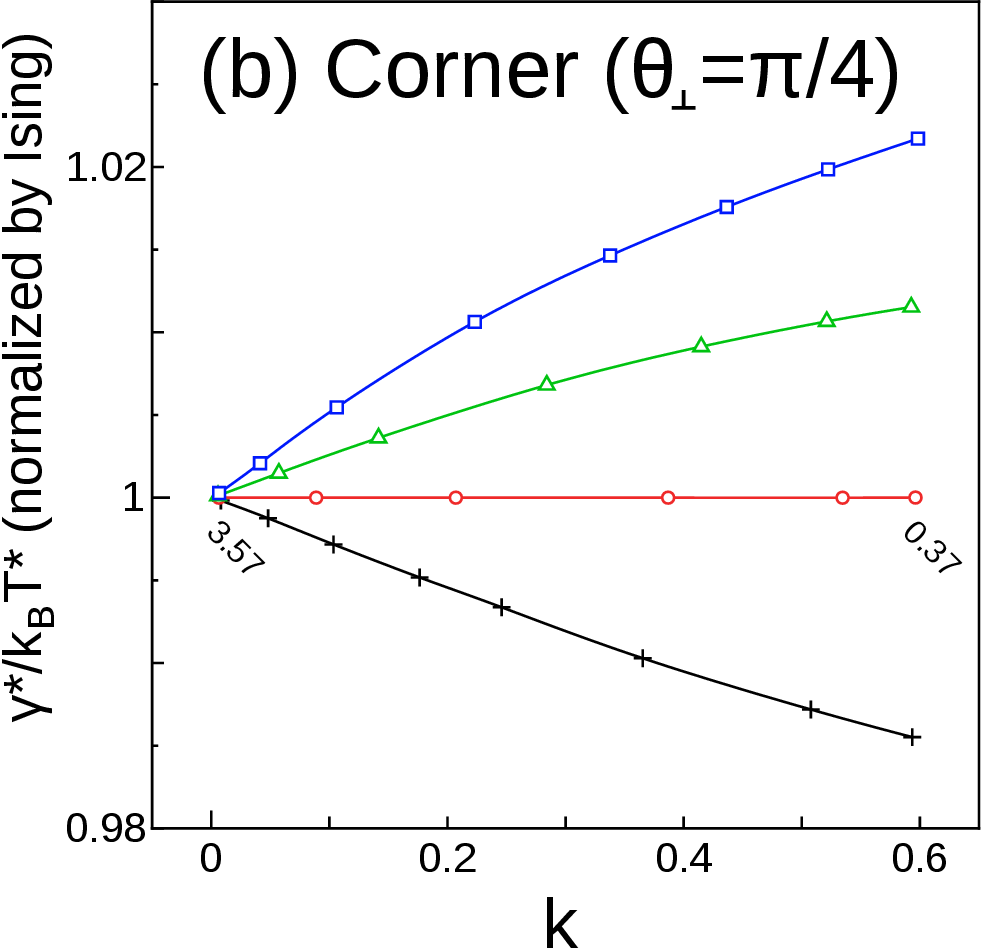}
   
   \vspace{3mm}
   \includegraphics[width=0.475\linewidth]{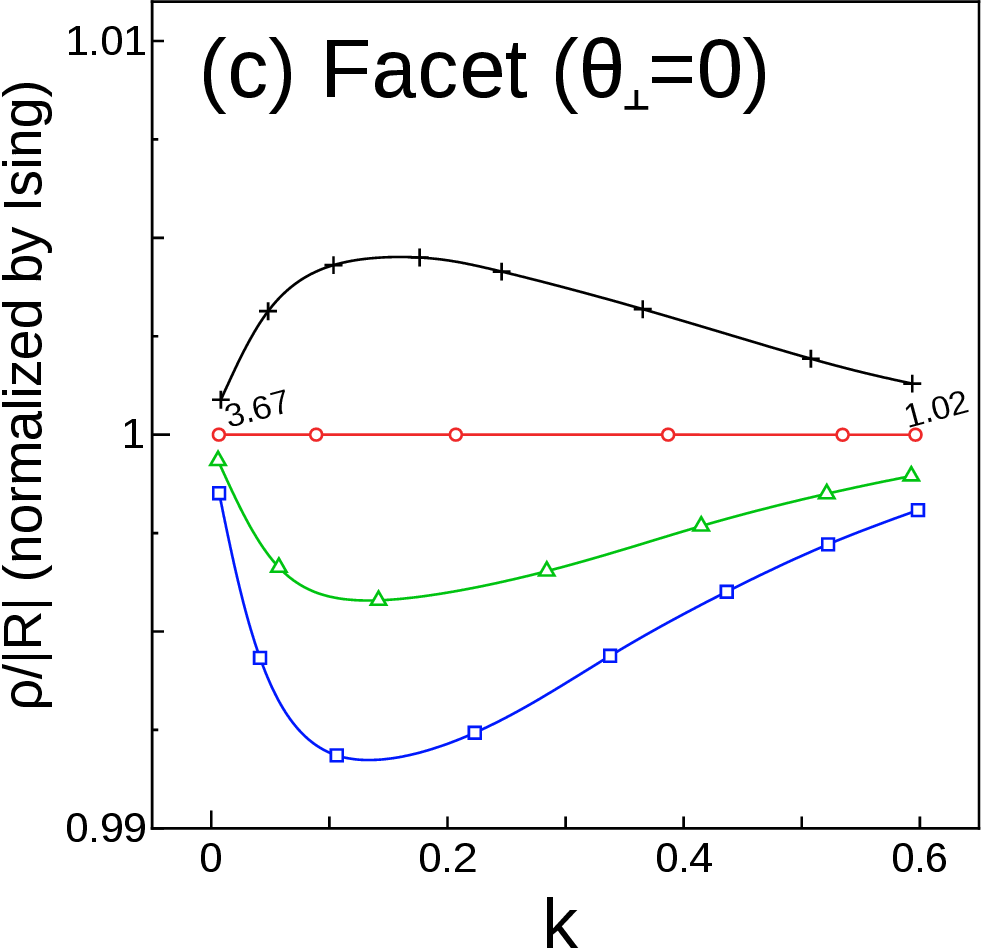}\hspace{2mm}
   \includegraphics[width=0.475\linewidth]{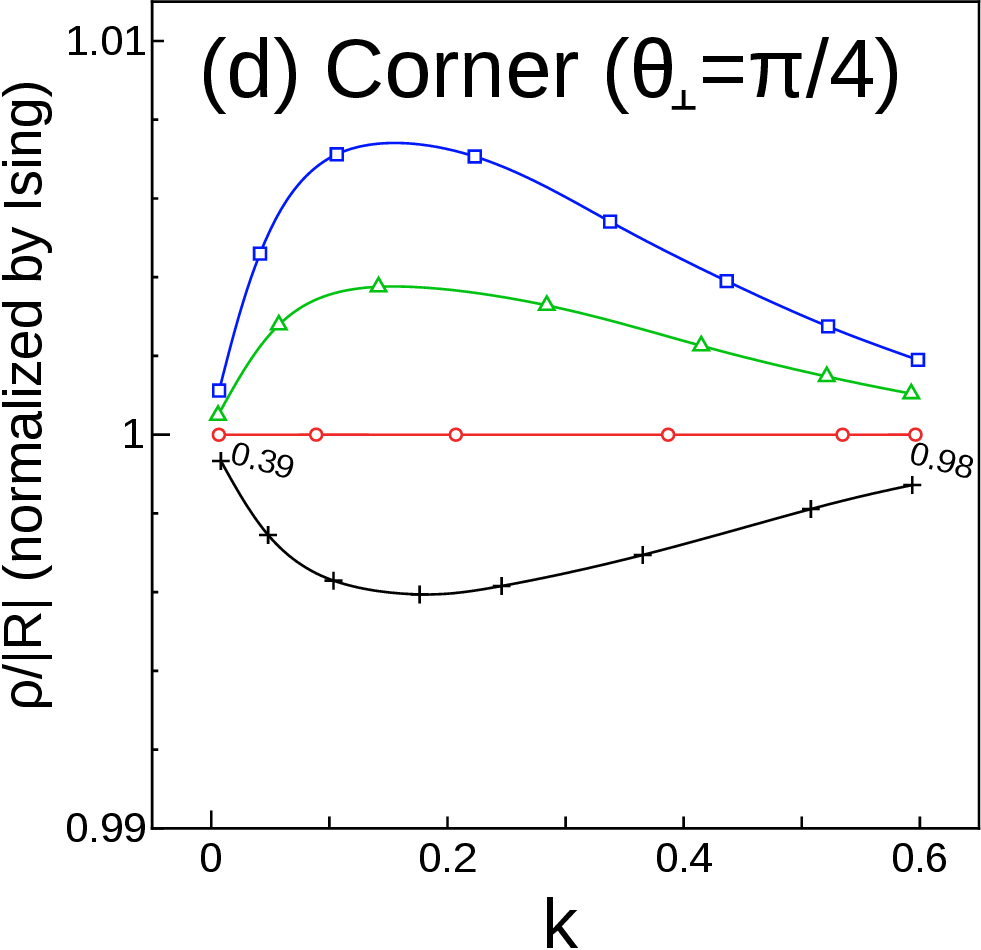}
  \end{center}
  \caption{%
  Elliptic modulus
  dependence of interfacial tension
  and radius of curvature in the facet ($\theta_\perp=0$)
  and the corner ($\theta_\perp=\frac\pi4$) directions.
  The correspondence between marks and $Q$ is provided in (a);
  the fitting curves are a guide for the eye.
  The figures plot the data normalized by the corresponding values
  of $Q=2$, which are calculated for an extracted $\bar k$ by taking
  $\bar b=1$ in \myEqs{eq:4.1} and (\ref{eq:4.5}).
  The bare Ising values are also given by numerals
  (see \myTable{table:2}).
  }
  \label{fig:3}
 \end{figure}
 }

 \newcommand\FIGAone{%
 \begin{figure}[t]
  \begin{center}
   \includegraphics[width=0.95\linewidth]{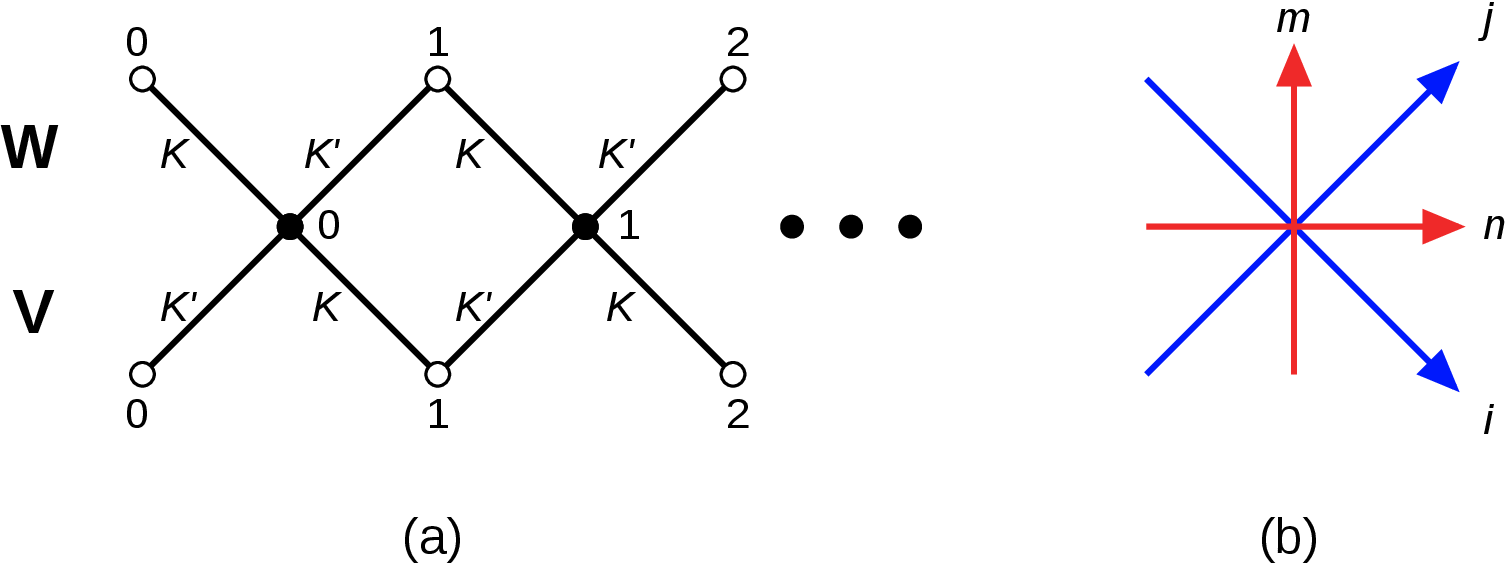}
  \end{center}
  \caption{%
  (a) Three successive rows of the square lattice drawn diagonally. 
  (b) The transfer matrix ${\bf U}$ operates spins on a row to
  transfer them from lower to upper direction along the
  $m$-axis.
  The $i$ and $j$ axes correspond to  the directions of 
  the primitive translation vectors ${\bf e}_x$ and ${\bf e}_y$
  of the square lattice,
  respectively.
  }
  \label{fig:A1}
 \end{figure}
 }
 
 \newcommand\FIGAtwo{%
 \begin{figure}[t]
  \begin{center}
   \includegraphics[width=0.8\linewidth]{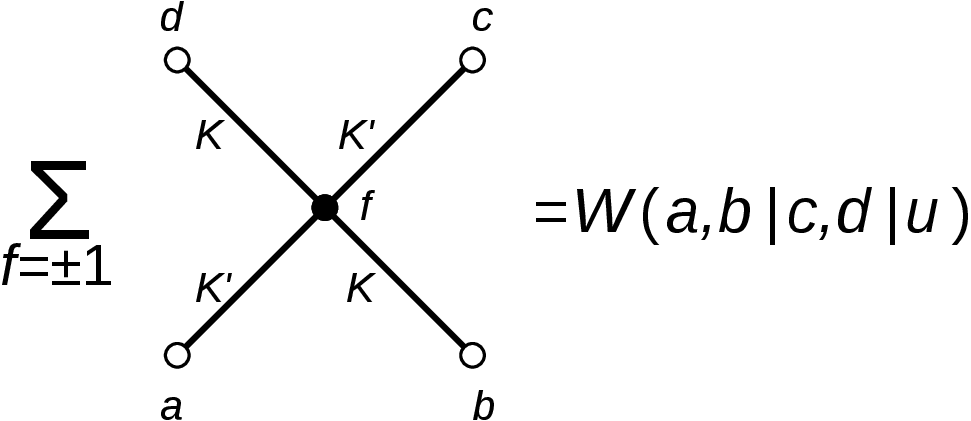}
  \end{center}
  \caption{%
  The Boltzmann weight of four edges $W(a,b|c,d|u)$,
  where $a$, $b$, $c$, and $d$ are the nearest-neighbor spins of $f$
  summed.
  }
  \label{fig:A2}
 \end{figure}
 }
 
 \newcommand\FIGBone{%
 \begin{figure}[t]
  \begin{center}
   \includegraphics[width=\linewidth]{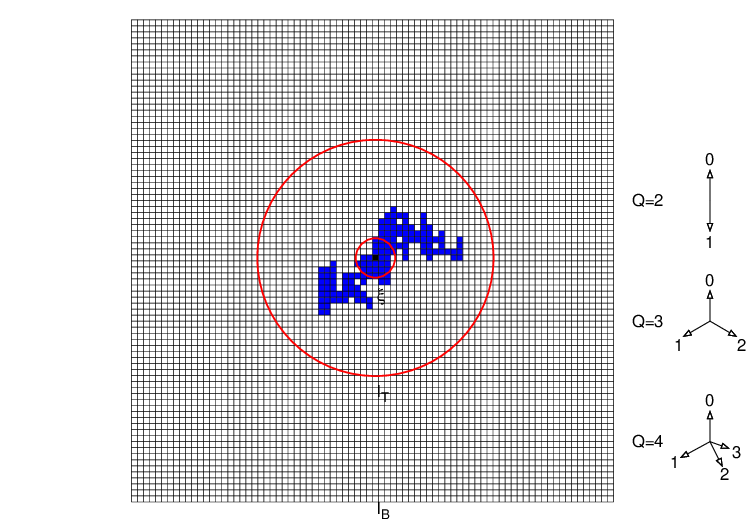}
  \end{center}
  \caption{%
  Left: Schematic representation of MC simulations.
  The black cell (the seed site) and blue cells exhibit the FK
  cluster ${\cal C}_i$.
  The length scales of the bounding box $l_{\rm B}$ and the
  equilibrated circular domain $l_{\rm T}$ as well as the
  correlation length $\xi$ are depicted.
  Right:
  We give the magnetic operators by $Q$ unit vectors in $Q-1$
  dimensions, where the corresponding value of $q_{\bf r}$ is
  denoted ($Q>1$).
  The arrows in the $Q=3$ ($Q=4$) case point to the corners of the
  regular triangle (tetrahedron).}
  \label{fig:B1}
 \end{figure}
 }
 
 \newcommand\FIGBtwo{%
 \begin{figure}[t]
  \begin{center}
   \includegraphics[width=0.475\linewidth]{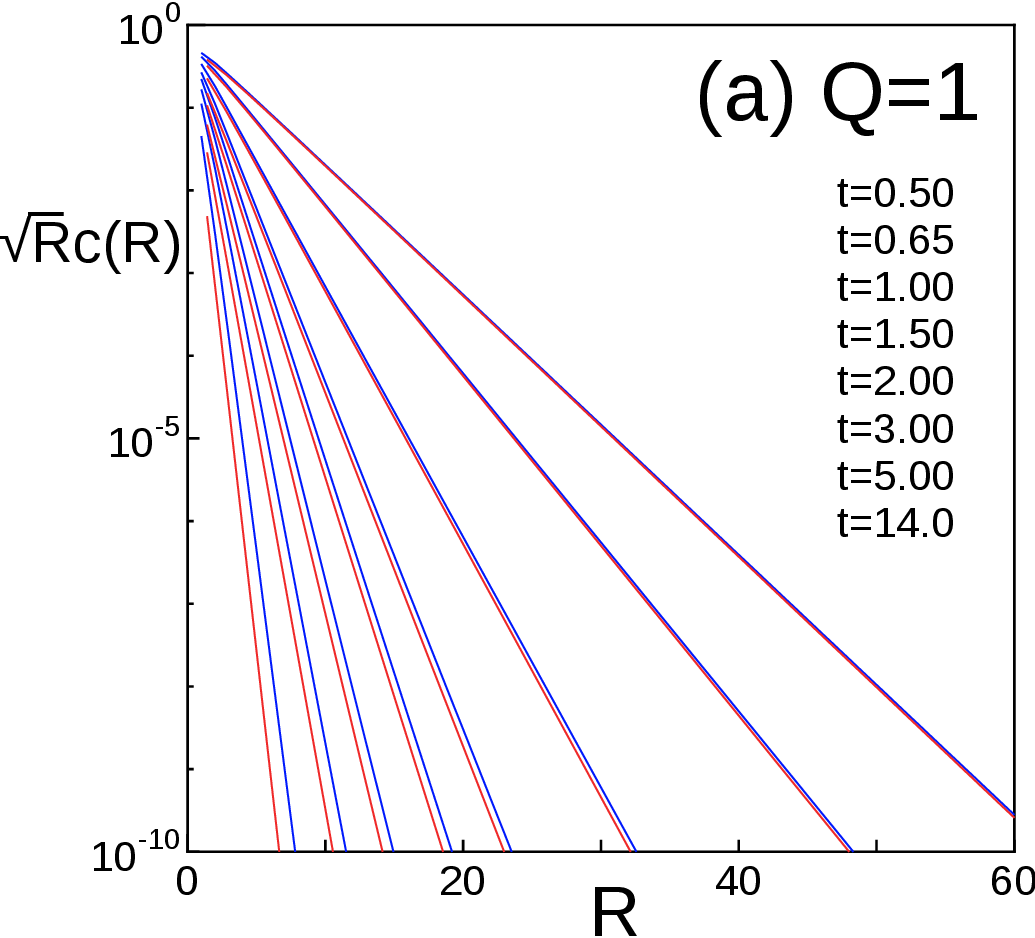}\hspace{2mm}
   \includegraphics[width=0.475\linewidth]{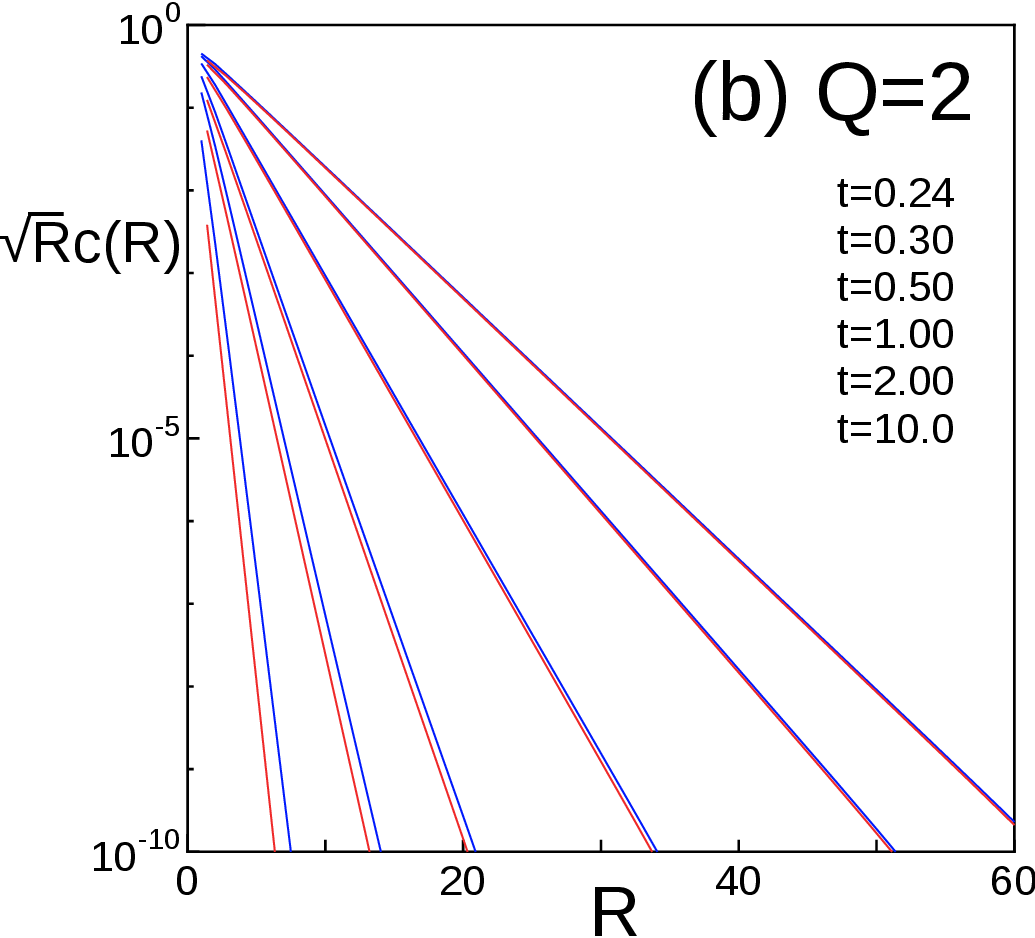}
   
   \vspace{3mm}
   \includegraphics[width=0.475\linewidth]{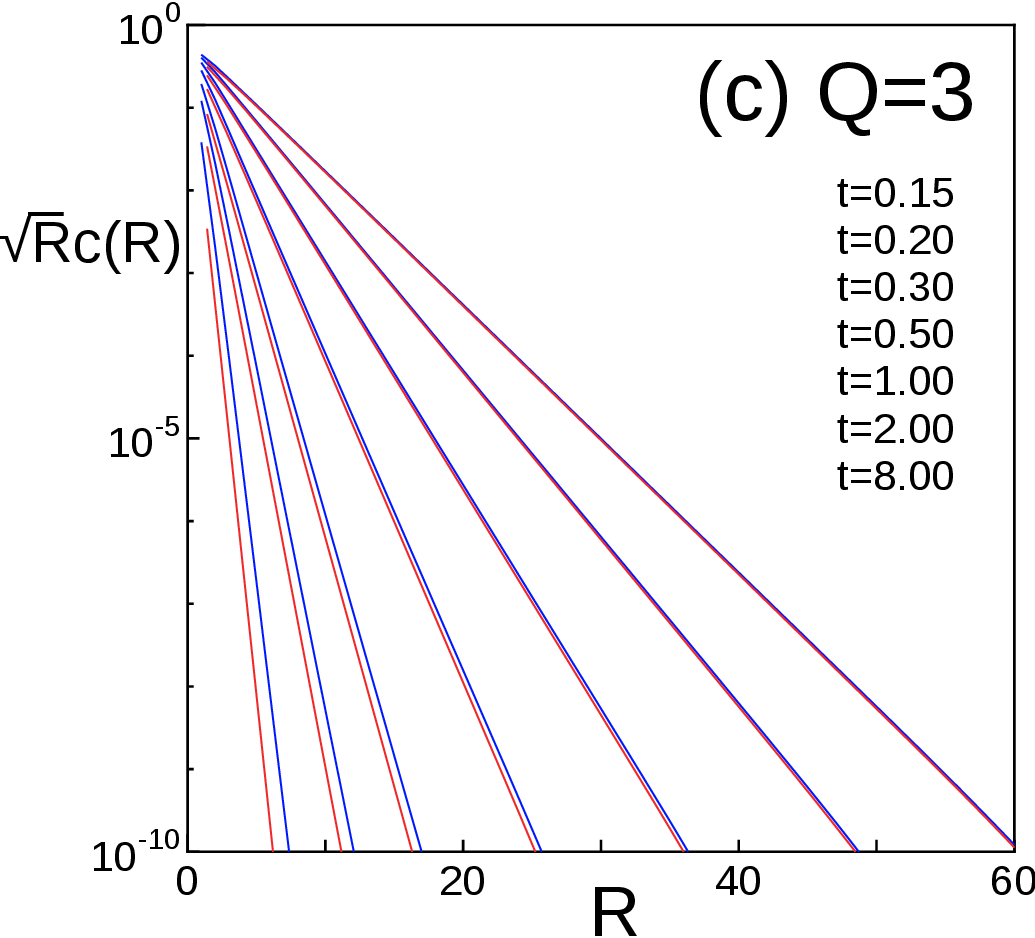}\hspace{2mm}
   \includegraphics[width=0.475\linewidth]{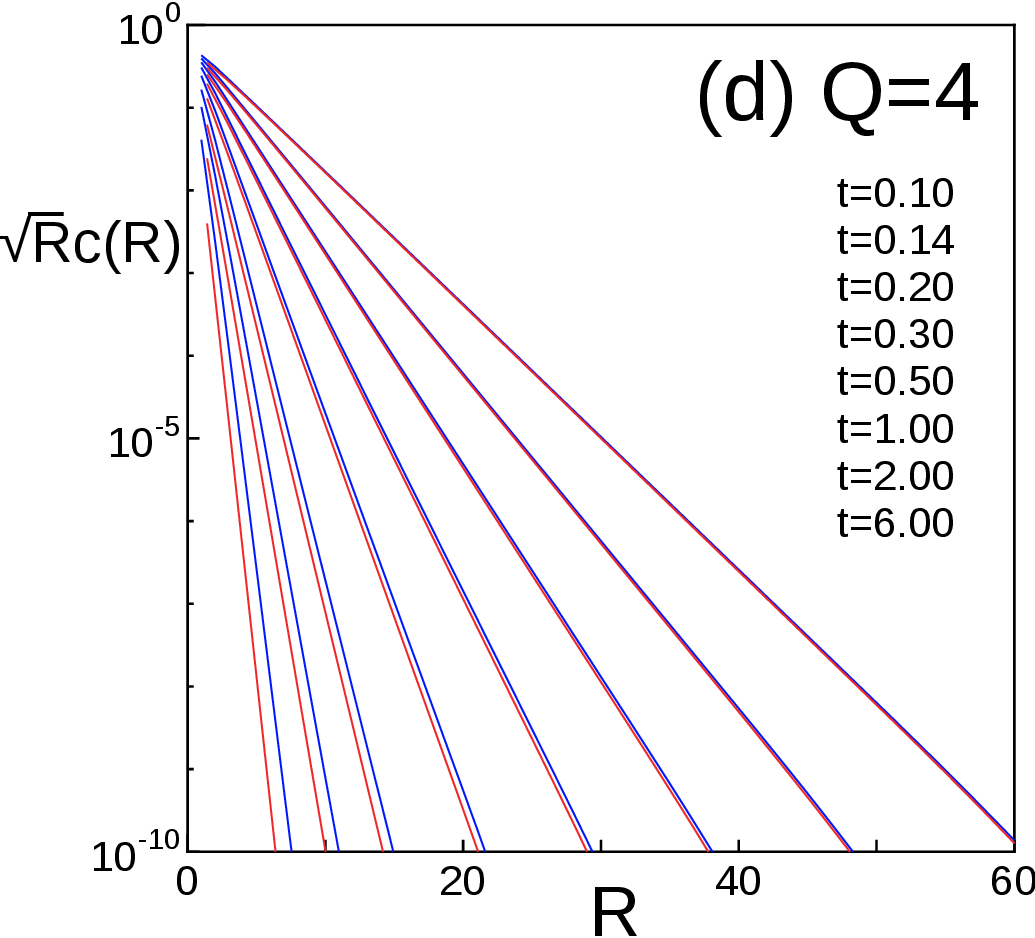}
  \end{center}
  \caption{%
  Correlation functions $c({\bf r})$ of the $Q$-state Potts
  model in two directions ($Q=1$ is the bond percolation).
  For clarity,
  we draw the lines for MC data points,
  where the blue (red) ones show $\sqrt{R}c(R)$
  in the row (diagonal) direction: 
  ${\bf r}=R{\bf e}_x$
  [${\bf r}=R({\bf e}_x+{\bf e}_y)/\sqrt{2}$].
  With the increase of the temperature $t$,
  the slopes of lines become steeper,
  and
  the discrepancies between blue and red lines become larger.}
  \label{fig:B2}
 \end{figure}
 }

 \newcommand\TABLEone{%
 \begin{table*}
  \caption{%
  \label{table:1}
  The temperature dependence of extracted values of the fitting
  parameters in \myEq{eq:3.2}.
  In addition, the correlation length $\xi_{\rm diag}$ is enumerated; 
  its exact values are given in the $Q=2$ case for
  comparison.
  The parenthesized digits indicate errors. 
  The geometries of annular regions employed for fittings are
  summarized (see text).}
  \begin{ruledtabular}
   \begin{tabular}{crccrlllll}
    $Q$&$t$~~~~~&$c_{\rm max}$&$c_{\rm min}$&$|{\myann}|$&$\bar A$&$\bar k$&$\bar b$&$\xi_{\rm diag}$&$\xi_{\rm exact}$\\
    \hline
    1
    & 0.50&$1\times10^{-4}$&$3\times10^{-5}$&436&1.024825(6)&0.593506     &1.018407(2)&2.750569(1)&n/a\\
    & 0.65&$1\times10^{-4}$&$2\times10^{-5}$&348&1.018603(5)&0.507724(4)  &1.01635(2) &2.113499   &\\
    & 1.00&$1\times10^{-4}$&$1\times10^{-5}$&252&1.01061(1) &0.365381(1)  &1.012548(5)&1.418417(1)&\\
    & 1.50&$1\times10^{-3}$&$1\times10^{-4}$&104&1.00565    &0.245847     &1.008724(3)&1.014686(1)&\\
    & 2.00&$1\times10^{-3}$&$1\times10^{-4}$& 68&1.00331(1) &0.176466     &1.006455   &0.819234(1)&\\
    & 3.00&$1\times10^{-3}$&$1\times10^{-4}$& 44&1.001351(1)&0.103562     &1.003900   &0.625435   &\\
    & 5.00&$1\times10^{-2}$&$1\times10^{-4}$& 36&1.000544(2)&0.0481419    &1.001796   &0.466765   &\\
    &14.00&$1\times10^{-2}$&$1\times10^{-5}$& 32&1.000040(5)&0.00817343(5)&1.000275   &0.294259   &\\
    \hline
    2
    & 0.24&$1\times10^{-3}$&$3\times10^{-4}$&308&1.000006(5)&0.596242(1)  &0.999987(7)&2.734823(4)&2.734823\\
    & 0.30&$1\times10^{-2}$&$1\times10^{-3}$&272&1.000000(3)&0.534544     &1.000006   &2.257906(2)&2.257906\\
    & 0.50&$1\times10^{-2}$&$1\times10^{-3}$&132&0.999998(2)&0.386861     &1.000001   &1.489133   &1.489133\\
    & 1.00&$1\times10^{-2}$&$1\times10^{-3}$& 52&0.999996(2)&0.207107     &1.000001(7)&0.898187   &0.898187\\
    & 2.00&$1\times10^{-1}$&$1\times10^{-3}$& 32&1.000001   &0.0888253    &1.000000   &0.584124   &0.584124\\
    &10.00&$1\times10^{-1}$&$1\times10^{-4}$& 20&0.999999   &0.00643374(2)&1.000000   &0.280253   &0.280253\\
    \hline
    3
    & 0.15&$1\times10^{-4}$&$3\times10^{-5}$&400&0.96856(4) &0.59271(1)   &0.98514(7) &2.672979(3) &n/a\\
    & 0.20&$1\times10^{-4}$&$3\times10^{-5}$&280&0.974811(4)&0.521168(7)  &0.98624(2) &2.147209(1) &\\
    & 0.30&$1\times10^{-4}$&$1\times10^{-5}$&284&0.982815(4)&0.414841(6)  &0.98813(2) &1.592772(2) &\\
    & 0.50&$1\times10^{-4}$&$1\times10^{-5}$&152&0.99072(3) &0.284116(2)  &0.991031(2)&1.116243(5) &\\
    & 1.00&$1\times10^{-3}$&$1\times10^{-4}$& 48&0.996832(2)&0.141663     &0.995079(2)&0.7210346(1)&\\
    & 2.00&$1\times10^{-3}$&$1\times10^{-5}$& 60&0.99923(1) &0.0572655(3) &0.997884   &0.493739    &\\
    & 8.00&$1\times10^{-2}$&$1\times10^{-5}$& 24&0.999977(3)&0.00576756(2)&0.999812   &0.2742787(2)&\\
    \hline
    4
    & 0.10&$3\times10^{-5}$&$1\times10^{-5}$&428&0.930993(6)&0.59836(1)   &0.97189(5) &2.695252(3) &n/a\\
    & 0.14&$3\times10^{-5}$&$1\times10^{-5}$&264&0.94516(6) &0.522324     &0.97424(1) &2.13511(2)  &\\   
    & 0.20&$1\times10^{-4}$&$1\times10^{-5}$&332&0.95917(3) &0.436457(2)  &0.977090(3)&1.67632(1)  &\\   
    & 0.30&$1\times10^{-4}$&$1\times10^{-5}$&204&0.97308(3) &0.337726(1)  &0.980765(2)&1.28399     &\\   
    & 0.50&$1\times10^{-4}$&$1\times10^{-5}$&100&0.98605(3) &0.223150(3)  &0.98580(1) &0.932945(2) &\\   
    & 1.00&$1\times10^{-4}$&$1\times10^{-5}$& 40&0.995700(8)&0.106277     &0.992435(1)&0.627439    &\\   
    & 2.00&$1\times10^{-3}$&$1\times10^{-5}$& 48&0.998968(2)&0.0413266    &0.996967   &0.4429263(1)&\\   
    & 6.00&$1\times10^{-2}$&$1\times10^{-5}$& 32&0.999930(4)&0.00668751   &0.999560   &0.2823396(1)&\\
   \end{tabular}
  \end{ruledtabular}
 \end{table*}
 }
 
 \newcommand\TABLEtwo{%
 \begin{table*}
  \caption{\label{table:2}%
  The elliptic modulus
  dependence of the interfacial tension
  (\ref{eq:4.1}) and the radius of the curvature (\ref{eq:4.5}) in
  the facet ($\theta_\perp=0$) and the corner ($\theta_\perp=\frac\pi4$)
  directions.
  The parenthesized digits indicate errors. 
  $t^\ast$ denotes a reduced temperature related via the duality
  condition with $t$.}
  \begin{ruledtabular}
   \begin{tabular}{crllllll}
    &&&&\multicolumn{2}{c}{$\gamma^{\ast}/\mykB T^{\ast}$}&\multicolumn{2}{c}{$\rho/|{\bf R}|$}\\
    \cmidrule(r){5-6}\cmidrule(r){7-8}
    $Q$&$t$~~~~~&$t^{\ast}$&$\bar k$&$\theta_\perp=0$&$\theta_\perp=\frac{\pi}{4}$ 
    &$\theta_\perp=0$&$\theta_\perp=\frac\pi4$\\
    \hline
    1
    & 0.50&n/a     &0.593506     &0.362487    &0.363561    &1.024034   &0.976668   \\
    & 0.65&        &0.507724(4)  &0.470826    &0.473149    &1.040366(3)&0.961575(3)\\
    & 1.00&        &0.365381(1)  &0.697672    &0.705011    &1.088328(2)&0.920472(1)\\
    & 1.50&        &0.245847     &0.966719(1) &0.985527(1) &1.170490(2)&0.859568(1)\\
    & 2.00&        &0.176466     &1.186919(2) &1.220652(3) &1.260456   &0.803824   \\
    & 3.00&        &0.103562     &1.530284(5) &1.598887    &1.448106(2)&0.713801   \\
    & 5.00&        &0.0481419    &2.000909    &2.142406    &1.823115   &0.597195   \\
    &14.00&        &0.00817343(5)&3.007556(3) &3.398371(4) &3.361020(8)&0.409052   \\
    \hline                                                                       
    2                                                                            
    & 0.24&0.180449&0.596242(1)  & 0.364650   & 0.365654   &1.022308   &0.978298   \\
    & 0.30&0.212423&0.534544     & 0.441115   & 0.442888   &1.032748   &0.968541   \\
    & 0.50&0.296641&0.386861     & 0.665509   & 0.671532   &1.075470   &0.931052   \\
    & 1.00&0.423400&0.207107     & 1.087883   & 1.113353   &1.209256(1)&0.834354   \\
    & 2.00&0.542189&0.0888253    & 1.631399   & 1.711964   &1.506487   &0.691295   \\
    &10.00&0.726099&0.00643374(2)& 3.137726(1)& 3.568201(2)&3.666353(8)&0.391270   \\
    \hline
    3
    & 0.15&0.123678&0.59271(1)   & 0.373113   & 0.374116   &1.021751(7)&0.978826(6)\\
    & 0.20&0.155806&0.521168(7)  & 0.463792   & 0.465721   &1.033903(4)&0.967477(4)\\
    & 0.30&0.210524&0.414841(6)  & 0.623150   & 0.627836   &1.062267(6)&0.942242(5)\\
    & 0.50&0.293001&0.284116(2)  & 0.882588(4)& 0.895863(4)&1.129046(3)&0.888962(2)\\
    & 1.00&0.416299&0.141663     & 1.341439   & 1.386896   &1.319586(3)&0.772300(2)\\
    & 2.00&0.531350&0.0572655(3) & 1.903634(2)& 2.025360(3)&1.714394(4)&0.625274(1)\\
    & 8.00&0.692933&0.00576756(2)& 3.197038(3)& 3.645926(3)&3.815002(3)&0.383675   \\
    \hline
    4
    & 0.10&0.087335&0.59836(1)   & 0.370106   & 0.371023   &1.020045(5)&0.980445(5)\\
    & 0.14&0.116378&0.522324     & 0.466507(3)& 0.468359(3)&1.032331(2)&0.968928(1)\\
    & 0.20&0.155063&0.436457(2)  & 0.592715(4)& 0.596544(4)&1.053238   &0.950094   \\
    & 0.30&0.209177&0.337726(1)  & 0.770348(4)& 0.778821(4)&1.092567   &0.917077   \\
    & 0.50&0.290435&0.223150(3)  & 1.050345(4)& 1.071875(3)&1.18063(1) &0.85286(1) \\
    & 1.00&0.411331&0.106277     & 1.528807(2)& 1.593782(2)&1.419833(4)&0.725893(2)\\
    & 2.00&0.523800&0.0413266    & 2.100168   & 2.257712   &1.898554   &0.580533   \\
    & 6.00&0.656712&0.00668751   & 3.118181(2)& 3.541834(2)&3.609079(2)&0.394535   \\
   \end{tabular}
  \end{ruledtabular}
 \end{table*}
 }
 
 \newcommand\TABLECone{%
 \begin{table*}
  \caption{\label{table:C1}%
  The $\myann_{\alpha}$ dependence of the estimates of
  $\xi_{\rm diag}$ and the reduced $\chi^2$ values (see text).
  Other than
  the optimized region ($\myann_0$),
  one inner region ($\myann_{-1}$), and  
  two outer regions ($\myann_1$ and $\myann_2$) are defined by
  $c_{\rm max}$ and $c_{\rm min}$ at each $Q$ and $t$.
  $R_\alpha$ is the mean radius of annular region $\myann_{\alpha}$.}
  \begin{ruledtabular}
   \begin{tabular}{cclccccll}
    $Q$&$t$&$\myann_\alpha$&$c_{\rm max}$&$c_{\rm min}$
    &$|\myann_\alpha|$&$R_\alpha$&$\xi_{\rm diag}$&~~$\bar\chi^2_{|\myann_\alpha|}$\\
    \hline
    1
    &0.50&$\myann_{-1}$&$1\times10^{-2}$&$1\times10^{-3}$&432&11.9&2.750670(2)&865~~~~\\
    &    &$\myann_{ 0}$&$1\times10^{-4}$&$3\times10^{-5}$&436&22.1&2.750569(1)&~~~0.642\\
    &    &$\myann_{ 1}$&$1\times10^{-5}$&$4\times10^{-6}$&392&27.7&2.750564(4)&~~~0.329\\  
    &    &$\myann_{ 2}$&$5\times10^{-8}$&$3\times10^{-8}$&336&41.2&2.75066(7)  &~~~1.231\\
    \hline
    2
    &0.24&$\myann_{-1}$&$1\times10^{-2}$&$3\times10^{-3}$&188&10.3&2.734824(1)&~~~0.084\\
    &    &$\myann_{ 0}$&$1\times10^{-3}$&$3\times10^{-4}$&308&16.0&2.734823(4)&~~~0.499\\
    &    &$\myann_{ 1}$&$3\times10^{-5}$&$1\times10^{-5}$&452&24.9&2.734836(8) &~~~0.560\\
    &    &$\myann_{ 2}$&$1\times10^{-7}$&$5\times10^{-8}$&444&39.3&2.7348(1)  &~~~0.849\\
    \hline
    3
    &0.15&$\myann_{-1}$&$1\times10^{-2}$&$1\times10^{-3}$&400&11.5&2.672762(3)&240~~~~\\
    &    &$\myann_{ 0}$&$1\times10^{-4}$&$3\times10^{-5}$&400&21.3&2.672979(3)&~~~1.10\\
    &    &$\myann_{ 1}$&$2\times10^{-6}$&$1\times10^{-6}$&368&30.6&2.67291(5) &~~~0.810\\
    &    &$\myann_{ 2}$&$1\times10^{-7}$&$5\times10^{-8}$&436&38.3&2.67285(2) &~~~0.496\\
    \hline
    4
    &0.10&$\myann_{-1}$&$1\times10^{-2}$&$1\times10^{-3}$&400&11.5&2.694577   &758~~~~\\
    &    &$\myann_{ 0}$&$3\times10^{-5}$&$1\times10^{-5}$&428&24.3&2.695252(3)&~~~0.301\\
    &    &$\myann_{ 1}$&$2\times10^{-6}$&$1\times10^{-6}$&348&30.7&2.69525(1) &~~~0.114\\
    &    &$\myann_{ 2}$&$1\times10^{-7}$&$5\times10^{-8}$&452&38.5&2.6951(2)  &~~~0.223\\
   \end{tabular}
  \end{ruledtabular}
 \end{table*}
 }

\makeatletter
\renewcommand{\theequation}{%
  \arabic{section}.\arabic{equation}}
\@addtoreset{equation}{section}
\makeatother
 
\begin{document}
 
 \title{%
   Asymptotic correlation functions in the $\boldsymbol{Q}$-state Potts
   model:\\
   a universal form for point group $\boldsymbol{C_{4v}}$}
   \author{Masafumi Fujimoto}
   \affiliation{%
   Department of Physics, Nara Medical University, Kashihara, Nara
   634-8521, Japan}
   \author{Hiromi Otsuka}
   \affiliation{%
   Department of Physics, Tokyo Metropolitan University, Tokyo
   192-0397, Japan}
   
   \date{\today}
 
 \begin{abstract}
  Reexamining algebraic curves found in the eight-vertex model,
  we propose an asymptotic form of the correlation functions for
  off-critical systems possessing rotational and mirror symmetries 
  of the square lattice, i.e., the $C_{4v}$ symmetry.
  In comparison with the use of the Ornstein-Zernike form,
  it is efficient to investigate
  the correlation length with its directional dependence (or anisotropy).
  We investigate the $Q$-state Potts model on the square lattice.
  Monte Carlo (MC) simulations are performed using the infinite-size 
  algorithm by Evertz and von der Linden.
  Fitting the MC data with the asymptotic form above the critical 
  temperature, we reproduce the exact solution of
  the anisotropic correlation length (ACL) of the Ising model ($Q=2$) 
  within a five-digit accuracy.
  For $Q=3$ and 4,
  we obtain numerical evidence that the asymptotic form is
  applicable to their correlation functions and the ACLs.
  Furthermore, we successfully apply it to the bond percolation 
  problem which corresponds to the $Q\rightarrow1$ limit.
  From the calculated ACLs,
  the equilibrium crystal shapes (ECSs) are derived via duality and
  Wulff's construction.
  Regarding $Q$ as a continuous variable, we find that the ECS of the
  $Q$-state Potts model is essentially the same as those of
  the Ising models on the Union Jack and 4-8 lattices, which
  are represented in terms of a simple algebraic curve of genus~1.
 \end{abstract}
 
 
 \maketitle
 
 \section{INTRODUCTION}
 \label{sec:1}
 
 For the past few decades,
 thermal evolution of the equilibrium crystal shape (ECS)
 \cite{Wulff1901,Burton1951}
 has received considerable attention
 \cite{Rottman1981,Avron1982,Zia1982,Beijeren1977,Jayaprakash1983,
 Fujimoto1997,Fujimoto1992,Fujimoto1993}.
 This revived interest comes from connections between the ECS and the
 roughening transition phenomena
 \cite{Burton1951,Rottman1981,Avron1982,Zia1982,
 Beijeren1977,Jayaprakash1983}.
 The first exact analysis of the ECS was done for the square-lattice
 Ising model
 \cite{Rottman1981};
 see also
 \myRefs{Avron1982,Zia1982}.
 
 Here
 we investigate the square-lattice Potts model
 \cite{Baxter1982,Wu1982}.
 To each site ${\bf r}$
 one associates a $Q$-valued variable $q_{\bf r}$.
 The Hamiltonian is given by
 \begin{equation}
  E(Q)
   =
   -\sum_{\langle{\bf r},{\bf r}'\rangle}
   J_{{\bf r, r}'}
   \delta\left(q_{\bf r},q_{{\bf r}'}\right),\quad
   q_{\bf r}=0,1,\dots,Q-1,
   \label{eq:1.1}
 \end{equation}
 where the sum runs over all nearest-neighbor pairs
 $\langle{\bf r},{\bf r}'\rangle$.
 Note that the $Q=2$ Potts model is equivalent to the Ising model.
 For general $Q$, 
 the Potts model is exactly solvable at the phase transition point
 \cite{Baxter1982,Temperley1971,Baxter1973,Kluemper1989,
 Buffenoir1993,Borgs1992}.
 The phase transition is continuous for $Q\leq 4$ and first order for
 $Q>4$.
 
 In a previous study
 \cite{Fujimoto1997}
 we generalized the argument in
 \myRefs{Rottman1981,Avron1982,Zia1982}
 to find the ECS of the $Q$-state Potts model.
 We showed that the anisotropic correlation length (ACL) is related by
 duality
 \cite{Zia1978,Laanait1987,Holzer1990,Akutsu1990}
 to the anisotropic interfacial tension.
 For $Q>4$,
 the ACL was exactly calculated at the first-order transition (or
 self-dual) point.
 The ECS was obtained from the ACL via the duality relation and
 Wulff's construction
 \cite{Wulff1901,Burton1951}.
 It was expressed as an algebraic curve in the $\alpha\beta$ plane 
 \begin{equation}
  \alpha^2\beta^2+1+A_3(\alpha^2+\beta^2)+A_4\alpha\beta=0,
   \label{eq:1.2}
 \end{equation}
 where 
 $\alpha=\exp[-\lambda(X+Y)/\mykB T]$ and 
 $\beta =\exp[-\lambda(X-Y)/\mykB T]$ with the position vector $(X,Y)$
 of a point on the ECS and a suitable scale factor $\lambda$; 
 for definitions of $A_3$ and $A_4$, see
 \exSec{3.2} of \myRef{Fujimoto1997}.
 The algebraic curve (\ref{eq:1.2}) is quite universal because it
 appears as the ECSs of a wide class of lattice models
 including the square-lattice Ising model
 \cite{Rottman1981,Avron1982,Zia1982,Beijeren1977,Jayaprakash1983,
 Fujimoto1997,Fujimoto1992,Fujimoto1993}.
 
 We note that \myEq{eq:1.2} is not the only universal curve
 \cite{Rutkevich2002,Fujimoto2002}.
 For example,
 we considered the ACL of the eight-vertex model in
 \myRef{Fujimoto1996}
 and found another algebraic curve
 \begin{equation}
  \alpha^2\beta^2
   +
   1
   +
   {\bar A}_2(\alpha\beta+1)(\alpha+\beta)
   +
   {\bar A}_3(\alpha^2+\beta^2)
   +
   {\bar A}_4\alpha\beta
   =
   0
   \label{eq:1.3}
 \end{equation}
 [for definitions of
 ${\bar A}_2$, ${\bar A}_3$, and ${\bar A}_4$,
 see \exEq{(4.7)} of \myRef{Fujimoto1996}].
 The ACL represented by \myEq{eq:1.3} is indeed the same as those of
 the Ising models on the Union Jack and 4-8 lattices
 \cite{Holzer1990a}.
 Some authors derived algebraic curves for the lattice models
 possessing six-fold rotational symmetry
 \cite{Holzer1990a,Vaidya1976,Zia1986,
 Fujimoto1999,Fujimoto1998,Fujimoto2002a},
 which are also universal.

 We expect that these algebraic curves are connected with the
 symmetries of lattice models.
 How do they select the algebraic curves?
 This is the problem we shall consider.
 Also,
 we expect that the same selection mechanism works regardless of
 whether the lattice models are exactly solvable or not.

 In this paper,
 we propose an asymptotic form of the correlation functions of
 off-critical systems possessing rotational and mirror symmetries of the
 square lattice (the $C_{4v}$ symmetry); see \exchap{2} of
 \myRef{Hamermesh1989}.
 The asymptotic form is brought about by reexamining exact solutions
 of the eight-vertex model.
 We apply it to the $Q$-state Potts model.
 Our method is a combined use of the asymptotic form and Monte Carlo
 (MC) simulations based on the Fortuin-Kasteleyn random-cluster
 representation
 \cite{Fortuin1972}.
 As we see below,
 the combined method is quite efficient to calculate the correlation
 lengths with their anisotropy.
 
 The format of the present paper is as follows:
 In \mySec{sec:2},
 we introduce an asymptotic form for the $C_{4v}$ symmetry, i.e.,
 a form for the asymptotic
 correlation functions which together with MC data enables us to
 evaluate the ACLs.
 In \mySec{sec:3}, we perform MC simulations. 
 We investigate the $Q=2$, 3, and 4 cases, and also the bond percolation
 model which is, via the cluster representation, realized in the
 $Q\rightarrow1$ limit.
 \mySec{sec:4}
 is devoted to discussion and summary.
 From the evaluated ACL we derive the ECS in the $Q$-state Potts
 model.
 Detailed explanations on
 the exact calculation of the Ising model,
 the methodology of MC simulations,
 and
 the fitting procedure are given in the Appendices.

 \section{ASYMPTOTIC CORRELATION FUNCTIONS FOR $\boldsymbol{C_{4v}}$}
 \label{sec:2}
 
 Johnson, Krinsky, and McCoy (JKM)
 \cite{Johnson1973}
 calculated the correlation length of the eight-vertex model along the
 vertical direction; see also
 \myRef{Johnson1972}.
 Their approach was the row-to-row transfer matrix argument.
 They investigated the low-lying excitations to determine the
 next-largest and next-next-largest eigenvalues.
 In
 \myRef{Fujimoto1996},
 using the shift operator, we extended the analysis by JKM into
 general directions; see also
 \myRefs{Fujimoto1990,Fujimoto1990a,Kluemper1990}.

 Because of the symmetry properties of the model,
 we can restrict ourselves to an antiferroelectric ordered regime
 (the principal regime)
 without loss of generality
 \cite{Baxter1982,Baxter1973a}.
 It was shown that, for a given parameter $x$ ($0<x<1$),
 there are two cases with respect to another parameter $q$
 (see \myFig{fig:1}) \cite{Fujimoto1996};
 for definitions of $x$ and $q$,
 see \exchap{10} of \myRef{Baxter1982}.
 In the case $0<q<x^3$ the ACL is independent of $q$. 
 In the $q\rightarrow x^4$ limit the eight-vertex model factors into
 two square-lattice Ising models.
 For planar Ising models it was shown that the ECS is determined by
 the Fourier transform (structure factor) of the asymptotic
 correlation function
 \cite{Holzer1990,Akutsu1990,Holzer1990a}.
 In the square-lattice Ising model the inverse of the structure factor
 above the critical temperature corresponds to the left-hand side
 of \myEq{eq:1.2}.
 We found that for $0<q<x^3$ the asymptotic correlation function of 
 the eight-vertex model is related to the algebraic curve
 (\ref{eq:1.2}).
 In the case $x^3<q<x^2$ the ACL depends on $q$. 
 It was shown that the asymptotic correlation function  
 is connected with \myEq{eq:1.3}.

 \FIGone
 
 The correlation function in the square-lattice Ising model (with
 ferromagnetic couplings) were investigated by many authors;
 see, for example,
 \myRefs{Kadanoff1966,Cheng1967,McCoy1973,
 Yamada1983,Yamada1984,Yamada1986,Wu1976,Camp1971,Fisher1971,Bariev1975}.
 The Pfaffian method was used in
 \myRef{Cheng1967}; see also \myRef{McCoy1973}. 
 Yamada
 \cite{Yamada1983,Yamada1984}
 showed that the results in
 \myRef{Cheng1967}
 coincide with those of the row-to-row transfer matrix.
 We note that the row direction of the eight-vertex model
 corresponds to the diagonal direction in the Ising model;
 transfer matrices of the two models have complex eigenvalues.
 In the thermodynamic limit, due to their continuous distribution, the
 summation over the eigenvalues becomes contour integrals.
 JKM showed that analyticity of the integrand (or eigenvalues) plays
 an important role:
 To compare their results in the decoupling limit with those in
 \myRef{Cheng1967},
 JKM rewrote the latter by the use of elliptic functions, which
 connect the structure factor with the eigenvalues along the row
 direction.
 Then, using the analytic property,
 they shifted the integration paths suitably to find equivalence
 between the results along the row and diagonal directions;
 see \exEqs{(3.5)}~and~(3.6) in JKM.

 In
 \myRef{Fujimoto2002}
 we discussed a close relation between the $C_{4v}$ symmetry and the
 algebraic curves (\ref{eq:1.2}) and (\ref{eq:1.3}).
 The eight-vertex model was defined on a square lattice rotated through 
 an arbitrary angle with respect to the coordinate axes
 \cite{Fujimoto1994}.
 Calculating eigenvalues of transfer matrices along various
 directions, we showed that lattice rotations shift (or deform) the
 integration paths.
 We pointed out that, to derive the equivalence between the results by
 transfer matrices along various directions, two further properties
 are needed in addition to (i) the analytic property found by JKM:
 (ii) a functional equation corresponding to the $\pi$-rotational
 invariance, and
 (iii) doubly periodic structure. 
 We argued that the properties (i)--(iii) essentially determine the
 asymptotic form of the correlation function possessing the $C_{4v}$
 symmetry.
 
 To ensure the argument in
 \myRef{Fujimoto2002},
 and to show its applicability to unsolvable models,
 we consider the $Q$-state Potts model.
 Since the analysis in
 \myRef{Fujimoto2002}
 was about the correlation function between two arrow spins in the
 antiferroelectric ordered regime,
 some ambiguity remained to clarify the role of the $C_{4v}$ symmetry.
 We successfully applied the same argument as in
 \myRef{Fujimoto2002}
 to the square-lattice Ising model and then found that
 the properties (i)--(iii) are actually satisfied
 (see \myAppendix{appendix:A}).

 Regarding $Q$ as a continuous variable, we assume (i)--(iii).
 We estimate the leading asymptotic behavior of the correlation functions 
 for general $Q$ as follows
 (for clarity here we summarize the discussion given in
 \myAppendix{appendix:A6}):
 The property (iii) shows that, choosing a suitable parametrization,
 we can represent the asymptotic correlation function
 ${\cal F}_{{\bf o},{\bf r}}$
 as
 \begin{equation}
  {\cal F}_{{\bf o},{\bf r}}
   =
   \ {\rm Const.}
   \int^{\omega_1}_{-\omega_1}~d\Theta~
   {\cal Y}(\Theta)^j
   {\cal X}(\Theta)^i,
   \label{eq:2.1}
 \end{equation}
 with ${\bf r}=i{\bf e}_x+j{\bf e}_y$.
 ${\cal Y}(\Theta)$ corresponds to eigenvalues of the row-to-row
 transfer matrix, and
 ${\cal X}(\Theta)$ those of the shift operator.
 Both
 ${\cal X}(\Theta)$ and
 ${\cal Y}(\Theta)$ are doubly periodic functions;
 see \myAppendix{appendix:A6}.
 
 In the case $Q=2$ the $\pi$ rotation of the lattice corresponds to
 shifting the integration path by $\omega_2$.
 The property (ii) suggests the relations 
 ${\cal Y}(\Theta+\omega_2)
 ={\cal Y}(\Theta)^{-1}$
 and 
 ${\cal X}(\Theta+\omega_2)
 ={\cal X}(\Theta)^{-1}$.
 The property (i) indicates analyticity of
 ${\cal Y}(\Theta)$ and ${\cal X}(\Theta)$.
 It follows that
 ${\cal Y}(\Theta)$ and ${\cal X}(\Theta)$ must be of the form
 \begin{equation}
  {\cal Y}(\Theta)
   =
   \prod_{l=1}^{\nu } k^{\frac12}\mysn(\Theta+\alpha_l),
   \quad
   {\cal X}(\Theta)
   =
   \prod_{l=1}^{\nu'} k^{\frac12}\mysn(\Theta+v+\beta_l),
   \label{eq:2.2}
 \end{equation}
 where $k$ is the modulus corresponding to the modular parameter 
 $\tau=\omega_2/\omega_1$.
 When the interactions do not depend on bond directions, the Potts model
 possesses the fourfold rotational symmetry.
 We can set $v=\pm\omega_2/2$, $\nu=\nu'$, and $\alpha_l=\beta_l$.
 Since the correlation function is real valued,
 we find that $\tau$ must be purely imaginary, which ensures the
 $C_{4v}$ symmetry of the system as well.
 
 It follows from the case $Q=2$ that
 the simplest form with $\nu=2$ corresponding to the next-largest
 eigenvalues appears above $\myTc(Q)$,
 where we denote the phase transition temperature by $\myTc(Q)$,
 regarding it as a function of $Q$. 
 For parameters $\alpha_1$ and $\alpha_2$, we find two possibilities:
 $(\alpha_1-\alpha_2)/\omega_1$ is purely imaginary or a real number.
 Since $\alpha_1-\alpha_2=\omega_2/2$ in the case $Q=2$,
 we assume it is a pure imaginary number.
 As a result, we obtain for $T>\myTc(Q)$
 \begin{widetext}
 \begin{equation}
  {\cal F}_{{\bf o}, {\bf r}}
   =
   {\rm Const.}
   \int^{\omega_1}_{-\omega_1}~d\Theta~
   \left[
    k
    \mysn(\Theta+B)
    \mysn(\Theta-B)
   \right]^j
   \left[
    k
    \mysn(\Theta+\omega_2/2+B)
    \mysn(\Theta+\omega_2/2-B)
   \right]^i.
   \label{eq:2.3}
 \end{equation}
 \end{widetext}
 Since we cannot determine $\alpha_1-\alpha_2$ to be $\omega_2/2$
 solely
 from the $C_{4v}$ symmetry, we have introduced a parameter $B$.
 The structure factor of \myEq{eq:2.3} is related to the algebraic
 curve (\ref{eq:1.3});
 we can regard \myEq{eq:1.2} is a special limit of \myEq{eq:1.3}.
 The asymptotic form (\ref{eq:2.3}) is expected to be one of general
 forms for systems possessing the $C_{4v}$ symmetry.
 The algebraic curve (\ref{eq:1.3}) is an elliptic curve, i.e., an
 algebraic curve of genus 1.
 \myEq{eq:2.3} is a differential form on the elliptic curve
 (\ref{eq:1.3})
 \cite{Namba1984}.

 As mentioned at the end of \myAppendix{appendix:A6},
 for $T<\myTc(Q)$
 almost the same argument holds: 
 From the case $Q=2$, it follows that
 $\nu=4$,
 $\alpha_1-\alpha_2=\alpha_3-\alpha_4=\omega_2/2$, and
 $\alpha_1-\alpha_3$ is a real number.
 The only difference from the
 case of $T>\myTc(Q)$ is expected to be that two elliptic curves are
 needed to represent the asymptotic correlation function.
 
 \section{NUMERICAL ANALYSES FOR $\boldsymbol{Q}$-STATE POTTS MODEL}
 \label{sec:3}

 Following the results in \mySec{sec:2},
 we investigate the asymptotic correlation functions in the $Q$-state
 Potts model.
 For $Q\geq3$ the phenomena in three or more phase systems have been
 attracted much attention
 \cite{Selke1982,Peczak1989,Gupta1992};
 see also
 \myRefs{Kluemper1989,Buffenoir1993,Borgs1992,Janke1994,Janke1997a}.
 In
 \myRef{Selke1982}
 interface properties in $Q=3$ were studied. 
 As mentioned in \mySec{sec:1}, at the first-order transition point,
 the Potts model possesses the same ACL and ECS as the Ising model
 \cite{Fujimoto1997}.
 Although for $Q\geq 3$ we expect a deviation of the ECS from the
 $Q=2$ case,
 definite results on this subject have not been obtained yet;
 see, for example,
 \myRefs{Akutsu1987,Akutsu1987a,Selke1989}.

 Also,
 numerical calculations of the correlation functions and the
 correlation lengths have been frequently performed.
 One typical way of doing them is to analyze the exponential decay of
 correlation function data in a certain fixed direction provided by MC
 simulations on finite-size systems
 \cite{Janke1994,Janke1997a}.
 However,
 it is recognized that such an approach cannot always give an accurate
 estimation of the correlation lengths.
 Further,
 the present analysis on the ACLs is expected to become more difficult
 because of the following reasons:
 First, 
 the finite-size effects in the MC data can affect the analysis of
 anisotropy in an unexpected manner.
 Second,
 because the patterns of the ACLs observed in different models, but
 sharing the $C_{4v}$ symmetry, are similar to each other
 \cite{Akutsu1987,Akutsu1987a},
 they possess only slight differences (see also \myAppendix{appendix:B}).

 In this situation,
 we employ an algorithm for the MC simulations proposed by Evertz and
 von der Linden
 \cite{Evertz2001}.
 Since it is a method for infinite-size systems,
 an extrapolation of data to the thermodynamic limit is not
 necessary.
 For a system in a disordered phase,
 we generate the random clusters by taking the center site as a seed
 site and expand the thermally equilibrated area outward
 (see \myAppendix{appendix:C}).
 Then,
 we measure the correlation functions in the area well equilibrated,
 which are free of the finite-size effects. 
 Typically $O(10^{14})$ random clusters are generated at each
 temperature to attain the high accuracy of MC data
 \cite{Fortuin1972}.

 Our method is a combined use of the MC data with the result in
 \mySec{sec:2}, which is expected to be quite efficient to investigate
 the ACLs.
 In \mySec{sec:3.a},
 we shall introduce an asymptotic form for the correlation functions
 in the $Q$-state Potts model, which includes three parameters.
 To determine them, fitting calculations are performed for the MC
 data.
 In \mySec{sec:3.b},
 we calculate the ACL in the $Q=2$ Potts model to demonstrate an
 accuracy of our numerical analysis.
 In \mySec{sec:3.c},
 we investigate the cases $Q=3$, 4.
 In \mySec{sec:3.d}  
 we also apply the method to the bond percolation process
 corresponding to the $Q\rightarrow1$ limit.
 
 \subsection{A form for asymptotic correlation functions in
 $\boldsymbol{Q}$-state Potts model}
 \label{sec:3.a}

 As mentioned at the end of \mySec{sec:2},
 since the analysis for $T>\myTc(Q)$ is more fundamental,
 we shall restrict ourselves to the case $T>\myTc(Q)$ below.
 Suppose a square lattice $\Lambda_{\rm sq}$.
 We denote the position vector of a site on $\Lambda_{\rm sq}$ as
 ${\bf r}=i{\bf e}_x+j{\bf e}_y$.
 The definition of the correlation function for the $Q$-state Potts
 model is
 \begin{equation}
  \myav{\sigma_{\bf o}\sigma_{\bf r}^{\ast}}
   =
   \myav{\exp\left[\frac{2\pi\myIm(q_{\bf o}-q_{\bf r})}{Q}\right]}
   =
   \myav{\frac{Q\delta\left(q_{\bf o},q_{\bf r}\right)-1}{Q-1}},
   \label{eq:3.1}
 \end{equation}
 where $\sigma_{\bf r}=\exp(2\pi\myIm q_{\bf r}/Q)$.
 It was rigorously proven that the correlation function decays
 exponentially above the transition temperature and at the first-order
 transition point
 \cite{Wu1982,Laanait1987,Hintermann1978}.

 We shall concentrate on the case of $J_{{\bf r,r}'}=2J$, where the
 transition point is simply given by
 $\mykB\myTc(Q)/J=2/\ln\left(1+\sqrt{Q}\right)$.
 Based on the observation in \mySec{sec:2},
 we shall employ the inferred form in \myEq{eq:2.3}.
 The $C_{4v}$ symmetry permits the inclusion of one fitting
 parameter $B$ other than the elliptic modulus and a normalization
 factor.
 We take the Ising model in \myAppendix{appendix:A}
 as a reference.
 We replace $\omega_1$ and $\omega_2$ by $I$ and $I'$, respectively.
 In terms of elliptic functions,
 our form is rewritten as
\begin{widetext}
 \begin{equation}
  {\cal F}_{\rm sq}(i,j;A,k,b)
   =
   \frac{A}{\pi}(1-k^2)^{\frac14}
   \int^I_{-I}d\phi
   \left[
    k
    \mysn(\phi+b\frac{\myIm I'}{4})
    \mysn(\phi-b\frac{\myIm I'}{4})
   \right]^j
   \left[
    k
    \mysn(\phi+(2+b)\frac{\myIm I'}{4})
    \mysn(\phi+(2-b)\frac{\myIm I'}{4})
   \right]^i,
   \label{eq:3.2}
 \end{equation}
\end{widetext}
 where $b$ is introduced by $B=b\myIm I'/4$,
 and the normalization factor is represented using a parameter $A$;
 these refer to the exact values $A=1$ and $b=1$ in the $Q=2$ case.

 We denote MC data of \myEq{eq:3.1} with
 ${\bf r}=i{\bf e}_x+j{\bf e}_y\in\Lambda_{\rm sq}$ by $\{c(i,j)\}$.
 In an asymptotic region of large $R$,
 we perform a fitting of MC data to determine the three parameters
 $A$,~$k$,~$b$.
 Using the extracted values ${\bar A},~{\bar k},~{\bar b}$,
 the asymptotic correlation function is represented as
 \begin{equation}
  \myav{\sigma_{\bf o}\sigma_{\bf r}^{\ast}}
   \sim
   {\cal F}_{\rm sq}(i,j;{\bar A},{\bar k},{\bar b}).
   \label{eq:3.3}
 \end{equation}
 We can find the ACL from \myEq{eq:3.3} by the method of steepest
 descent, as shown in \myAppendix{appendix:A}.
 For example, the inverse correlation length in the diagonal direction is
 determined as
 \begin{equation}
  \frac{1}{\xi_{\rm diag}}
   =
   -\frac{1}{\sqrt 2}
   \ln
   \left\{
    \frac{\mysn[I+(1+{\bar b})\frac{\myIm I'}{4}]}
    {\mysn[I+(3+{\bar b})\frac{\myIm I'}{4}]}
    \frac{\mysn[I+(1-{\bar b})\frac{\myIm I'}{4}]}
    {\mysn[I+(3-{\bar b})\frac{\myIm I'}{4}]}
   \right\}.
   \label{eq:3.4}
 \end{equation}
 Note that, when ${\bar b}=1$,
 \myEq{eq:3.4} reduces to
 \begin{equation}
  \frac{1}{\xi_{\rm diag}}
   =
   -\frac{1}{\sqrt 2}
   \ln {\bar k},
   \label{eq:3.5}
 \end{equation}
 which coincides with the exact result in $Q=2$.
 If we succeed in calculating the ACL with a sufficient accuracy,
 then it gives strong numerical evidence that
 the elliptic curve~(\ref{eq:1.3}) appears in the structure factor of
 the asymptotic correlation function.
 
 There are two possible sources of errors in our analysis:
 One is the statistical errors in $\{c(i,j)\}$,
 which are inherent in the MC sampling procedures and become larger
 for longer distances.
 The other is systematic errors in \myEq{eq:3.2}.
 Note that contributions from the eigenvalues with $\nu>2$ are not
 taken into account in \myEq{eq:3.2}.
 They are small corrections to the asymptotic form, but can be
 important for short distances;
 see \myAppendix{appendix:A};
 although the accuracy of MC data is higher for shorter distances, the
 fitting results can be worse due to the systematic errors.
 
 We point out that essentially the same situations occur in methods
 along fixed directions and that these methods are not efficient to
 control the two kinds of errors;
 see, for example,
 \myRefs{Janke1994,Janke1997a}. 
 On the other hand, for the analysis of ACLs,
 it is rather natural to fit the MC data in an annular region.
 We do this with the help of \myEq{eq:3.2}.
 We found that, by optimizing a mean radius of the annular region, we
 can obtain reliable fitting results under a well-controlled condition
 of two kinds of errors.
 We provide details of our fittings below and in
 \myAppendix{appendix:C}.
 
 \FIGtwo
 
 \subsection{$\boldsymbol{Q=2}$ case}
 \label{sec:3.b}

 We start with the $Q=2$ Potts model and demonstrate an accuracy of
 our numerical analysis.
 We performed extensive MC simulations to achieve a demanded accuracy 
 and fitting the MC data in a suitable annular region.
 To make it explicit,
 let us denote an annular region centered at the origin as
 ${\myann}(c_{\rm max},c_{\rm min})=
 \{(i,j) | c_{\rm min}<c(i,j)<c_{\rm max}\}$,
 and the number of included sites as
 $|{\myann}(c_{\rm max},c_{\rm min})|$.
 For instance,
 at the reduced temperature
 $t=[T-T_{\rm C}(2)]/T_{\rm C}(2)=0.24$,
 we employed
 $|{\myann}(10^{-3},3\times10^{-4})|=308$
 with a mean radius $\simeq 16$,
 as given by blue cells in \myFig{fig:2}.
 The second column of \myTable{table:1} summarizes the results of $Q=2$.
 Then, one can find that, at all temperatures $t$,
 our results coincide with the exact values,
 $\xi_{\rm exact}$ along the diagonal direction,
 $A=1$,
 and
 $b=1$
 within, at least, five-digit accuracy.
 
 As explained in \myAppendix{appendix:A},
 the systematic errors for the $Q=2$ Potts model stem from the
 eigenvalues with $\nu=6$ and $r=-1$, which form the third band,
 and thus should be smaller than those in other cases.
 This permits us to use \myEq{eq:3.2} for inner annuli.
 We have checked a very weak dependence of fitting results on radii of
 inner annuli (see \myAppendix{appendix:C}).
 In outer regions the statistical errors become larger.
 However,
 we have also checked that their accuracy is improved by increasing
 the MC steps.
 If we increase the MC steps further,
 then the same results are expected to appear in outer annuli.
 Thus, as mentioned at the end of previous section, we can successfully 
 control the two kind of errors, which is the main advantage in our method 
 over calculations based on the Ornstein-Zernike form.

 Also, see the second column of \myTable{table:2} and the red lines in
 \myFig{fig:3}; we can confirm that \myEq{eq:3.2} is a quite efficient
 form of the asymptotic correlation functions, especially, to analyze
 the correlation lengths with their full anisotropies.
 
 \TABLEone

 \subsection{$\boldsymbol{Q=3,~4}$ cases}
 \label{sec:3.c}
 
 In this subsection,
 we analyze the ACLs observed in the $Q=3$, 4 Potts models by using
 the same method as in \mySec{sec:3.b}.

 As mentioned in \myAppendix{appendix:A},
 the contribution from the next-next largest eigenvalues with $\nu=4$
 vanishes due to the ${\mathbb Z}_2$ symmetry of the $Q=2$ Potts model.
 We cannot expect the same here.
 In fact,
 if we compare the $Q=2$ and $t=0.24$ case with the $Q=3$ and $t=0.15$
 case, though values for $\xi_{\rm diag}$ are nearly equal to
 each other
 (i.e., $\bar\xi\simeq2.7$),
 we cannot attain the same accuracy of the fitting for the latter MC
 data in the former annulus (i.e., blue cells in \myFig{fig:2}).
 We cannot attribute it to the statistical errors, but to an influence
 from the contribution of the next-next largest eigenvalues in the
 latter.
 Therefore,
 to circumvent the systematic errors,
 we need to employ annuli with larger radii than those in the $Q=2$
 case.
 For this issue,
 we estimate the order of errors included in \myEq{eq:3.2} to optimize
 the annulus employed in the $Q=2$ case.
 In the $Q=3$ and $t=0.15$ case,
 denote the deviation at the origin $(0,0)$ as $\Delta(3)$,
 which is estimated as $\Delta(3)\simeq O(10^{-2})$
 (see \myAppendix{appendix:C}).
 Noting that the next-next largest eigenvalues form the second band,
 we can estimate their contribution as
 $\Delta(3)\times{\rm e}^{-2R/\bar\xi}$.
 To obtain the ACL within a sufficient accuracy, we employ
 $|{\myann}(10^{-4},3\times10^{-5})|=400$ with the mean radius $\simeq 21$,
 which is depicted by the green cells in \myFig{fig:2}.
 The third column of \myTable{table:1} summarizes the results for the $Q=3$
 Potts model.
 We succeeded in fitting the data by \myEq{eq:3.2}, which permits us
 to evaluate the ACL of the model within five-digit accuracy at all
 temperatures calculated.
 As in the $Q=2$ case,
 we checked that the improvement of the accuracy was observed for the
 fitting of data in the outer annuli by increasing MC steps.
 The obtained results exhibit the relevant deviation from the values
 of the Ising models.
 
 In the $Q=4$ Potts model,
 the second band contributions become somewhat larger than those in
 the $Q=3$ case.
 This can be recognized via the same argument as above:
 We compare the $Q=3$ and $t=0.15$ case with the $Q=4$ and $t=0.10$ case;
 the correlation lengths in these cases are nearly equal. 
 We can also estimate the deviation $\Delta(4)$ and then find that it
 becomes larger than $\Delta(3)$.
 Therefore,
 we should employ a slightly larger annulus in radius than
 corresponding one in the $Q=3$ case.
 Based on the same order-estimate of the second band contributions as
 the above, for instance for $t=0.10$, we employed
 $|{\myann}(3\times10^{-5},10^{-5})|=428$ with mean radius $\simeq24$,
 which is indicated by the red cells in \myFig{fig:2}.
 The fitting can be performed with the same accuracy as in the $Q=3$
 case, and the obtained results are summarized in the fourth
 column of \myTable{table:1}.
 The deviation from the Ising model becomes more prominent.
 Note that the parameter $b$ monotonically decreases with the increase
 of $Q$;
 we will discuss its physical meanings in \mySec{sec:4}.
 
 \subsection{Bond percolation process as $\boldsymbol{Q\rightarrow1}$ limit}
 \label{sec:3.d}

 The Potts model is related to a number of other problems
 in lattice statistics
 \cite{Baxter1982,Wu1982,Temperley1971}.
 These relations make it possible to explore their properties from
 known results on the Potts model or vice versa
 \cite{Nijs1979,Nightingale1980,Bloete1982}.
 The bond percolation provides a simple picture of a phase transition
 \cite{Aharony2003}. 
 Regarding $Q$ as a continuous variable
 \cite{Bloete1982},
 we can relate the $Q$-state Potts model to the bond percolation
 model:
 Suppose that $Z(Q)$ is the partition function of the $Q$-state Potts
 model, whose cluster representation is provided in
 \myAppendix{appendix:B}.
 Then, the generating function of the bond percolation is given by
 \begin{equation}
  \lim_{Q\rightarrow1}
   \frac{\partial}{\partial Q}\ln Z(Q),
   \label{eq:3.6}
 \end{equation}
 where the bond percolation probability is $p=1-{\rm e}^{-2K}$
 \cite{Baxter1982},
 and
 the percolation threshold $p_{\rm C}$ is $p$ at $\myTc(1)$.
 The connectivity function is defined by the probability that the
 origin ${\bf o}$ and the site ${\bf r}$ belong to the same cluster,
 and was proven that, for $p<p_{\rm C}$, it decays exponentially 
 as ${\bf r}$ becomes large
 \cite{Higuchi1988,Aizenman1987,Menshikov1986}.
 The correlation function (\ref{eq:3.1}) reduces to the connectivity
 function in the $Q\rightarrow 1$ limit
 \cite{Wu1982}.
 Therefore, 
 in this subsection,
 we investigate the connectivity function in the bond percolation, 
 equivalently the correlation function of the Potts model in the
 $Q\rightarrow 1$ limit, by using \myEq{eq:3.2}.
 
 The first column of \myTable{table:1} summarizes the fitting results for
 $Q=1$.
 Based on the same argument as above,
 we optimized the annular regions:
 for example for $t=0.50$, we employ
 $|{\myann}(10^{-4},3\times10^{-5})|=436$ with mean radius $\simeq22$,
 which is given by crosses in \myFig{fig:2}.
 Then, we performed the fittings of the MC data in the optimized annuli 
 to determine $A,~k,~b$.
 We obtained the ACL within five-digit accuracy.
 In the course of fitting calculations, we recognized the systematic
 errors similarly to the $Q=3$, 4 cases.
 The extracted parameter values $\bar A$ and $\bar b$ exhibit deviations
 from the values of $Q=2$ in the opposite direction to the $Q=3$, 4
 cases and reveal their monotonic $Q$ dependence.
 
 As demonstrated,
 the form (\ref{eq:3.2}) can accurately fit the asymptotic correlation
 functions of the $Q=2$, 3, 4 Potts models for $T>\myTc(Q)$ and the
 asymptotic connectivity function of the bond percolation ($Q=1$) for
 $p<p_{\rm C}$.
 The result strongly suggests that
 the elliptic curve (\ref{eq:1.3}) is related to the structure factor
 for general $Q$.
 Consequently,
 we expect that \myEq{eq:1.3} plays a key role to describe the
 asymptotic correlation functions in a wide class of models
 possessing the $C_{4v}$ symmetry.
  
 \section{DISCUSSION AND SUMMARY}
 \label{sec:4}

 We have investigated the asymptotic correlation functions of the
 $Q$-state Potts model on the square lattice.
 Revisiting the exact solutions of the eight-vertex model,
 we pointed out the importance of the three properties of the eigenvalues 
 of the transfer matrix: 
 (i) the analyticity found by JKM
 \cite{Johnson1973}, 
 (ii) the functional equation related to the $\pi$-rotational invariance, 
 and
 (iii) the doubly periodic structure.
 Assuming (i)--(iii), we can essentially determine the asymptotic forms
 with the help of the $C_{4v}$ symmetry.

 For the off-critical Ising model $Q=2$ we proved that (i)--(iii) are
 satisfied;
 see \myAppendix{appendix:A}.
 For $Q>4$ the same situation occurs at the first-order transition
 point.
 Assuming $Q$ as a continuous variable,
 we brought the three properties into the asymptotic form with isotropic
 interactions above the transition temperature as \myEq{eq:3.2}.

 Based on these observations,
 we have proposed the new approach to analyze the correlation
 functions by using the result from (i)--(iii) and the numerical procedures
 combined:
 \myEq{eq:3.2} includes the parameters $A,~k,~b$ not determined from the
 $C_{4v}$ symmetry.
 We performed MC simulations provided by the infinite-size algorithm
 \cite{Evertz2001},
 and then carried out fittings of MC data to determine $A,~k,~b$.

 As mentioned in \mySec{sec:3}, there are two types of errors:
 statistical errors and systematic errors.
 One typical way of calculating the correlation 
 lengths is to consider the exponential decay of the correlation function 
 along fixed directions, but this method is not efficient to control the 
 two types of errors. We handled these errors successfully
 by introducing the annular regions for the fitting of $A,~k,~b$.

 To demonstrate the efficiency of our approach,
 we calculated the ACL of the Ising model above the critical
 temperature.
 The obtained results agreed extremely well with the exact values.
 We investigated the $Q=3$, 4 Potts models for $T>\myTc(Q)$ and then
 the bond percolation model (the $Q\to1$ limit) for $p<p_{\rm C}$.
 To minimize the errors, the annular regions were optimized carefully.
 We succeeded in fitting the data within a five-digit accuracy.
 The high accuracy of the results for $Q=1$, 2, 3, and 4 shows the 
 validity of the asymptotic form (\ref{eq:3.2})
 and that our approach is in fact effective to investigate the ACLs 
 of the system possessing the $C_{4v}$ symmetry.

 \subsection{Equilibrium crystal shapes}
 \label{sec:4.a}
 
 It was revealed that
 (I)
 the structure factors of the $Q$-state Potts model, 
 including the bond percolation as $Q=1$,
 are represented by the use of the elliptic curve (\ref{eq:1.3}) and
 that
 (II)
 the parameter $\bar b$ monotonically decreases with the
 increase of $Q$. 
 It is noticeable that, although small in magnitude, 
 (II) provides the reliable evidence of deviation from the case of the
 Ising model.
 Here, to show its physical meanings, 
 we investigate the $Q$ dependence of the ECS.

 The ECS is the droplet shape of one phase embedded inside a sea of
 another phase with its volume (or area) fixed
 \cite{Wulff1901,Burton1951,Rottman1981,Avron1982,Zia1982,
 Beijeren1977,Jayaprakash1983,Fujimoto1997,Fujimoto1992,Fujimoto1993}.
 Disappearance of facet in the ECS is a signal of the roughening
 transition
 \cite{Jayaprakash1983,Jayaprakash1984,Holzer1989}.
 Once knowing the anisotropic interfacial tension,
 we can determine the ECS with the help of Wulff's construction
 \cite{Wulff1901,Burton1951}.
 
 In a previous work
 \cite{Fujimoto1997},
 we found that for $Q\geq2$ the ACL is related to the anisotropic
 interfacial tension as
 \begin{equation}
  \frac{\gamma^{\ast}}{\mykB T^{\ast}}=\frac{1}{\xi}
   \quad
   {\rm in~all~directions},
   \label{eq:4.1}
 \end{equation}
 where $\gamma^{\ast}$ is the anisotropic interfacial tension at a
 temperature $T^{\ast}$ [$<\myTc(Q)$]
 such that $K^{\ast}=J/\mykB T^{\ast}$
 satisfies the duality condition
 $({\rm e}^{2K}-1)({\rm e}^{2K^{\ast}}-1)=Q$.
 We regard $\gamma^{\ast}$ as a function of $\theta_{\perp}$, 
 which is the angle between the normal vector of the interface and 
 ${\bf e}_x$-direction; $\theta=\theta_{\perp}+\pi/2$.
 The ECS is derived from $\gamma^{\ast}(\theta_{\perp})$ with the help of
 Wulff's construction as
 \begin{equation}
  \Lambda{\bf R}
   =
   \left(\!\!
    \begin{array}{rr}
     \cos\theta_{\perp}&-\sin\theta_{\perp}\\
     \sin\theta_{\perp}& \cos\theta_{\perp}
    \end{array}
    \!\!\right)
   \left(\!\!
    \begin{array}{c}
     \gamma^{\ast}(\theta_{\perp})\\
     \frac{d\gamma^{\ast}}{d\theta_{\perp}}(\theta_{\perp})
    \end{array}
    \!\!\right),
   \label{eq:4.2}
 \end{equation}
 where ${\bf R}=(X,Y)$ is the position vector of a point on the ECS and
 $\Lambda$ a scale factor adjusted to yield the area of the crystal.

 Using $\xi$ calculated in \mySec{sec:3} in \myEq{eq:4.1},
 we can derive the ECS via \myEq{eq:4.2}.
 Our result is as follows:
 \begin{equation}
  \Lambda{\bf R}
   =
   \left(\!\!
    \begin{array}{c}
     -\ln
      \left[
       {\bar k}
       \mysn(\phi+{\bar b}\frac{\myIm I'}{4})
       \mysn(\phi-{\bar b}\frac{\myIm I'}{4})
      \right]\\
     -\ln
      \left[
       {\bar k}
       \mysn(\phi-(2+{\bar b})\frac{\myIm I'}{4})
       \mysn(\phi-(2-{\bar b})\frac{\myIm I'}{4})
      \right]
    \end{array}
    \!\!\right)
   \label{eq:4.3}
 \end{equation}
 with $\Lambda$ chosen suitably.
 As $\phi$ moves from 0 to $2\myIm I'$ on the imaginary axis, ${\bf R}$
 sweeps out the ECS.
 One finds that, reflecting the result (I), the ECS is expressed as
 \myEq{eq:1.3} with
 $\alpha=\exp(-\Lambda X)$ and
 $\beta =\exp(-\Lambda Y)$,
 where
 ${\bar A}_2$, ${\bar A}_3$, ${\bar A}_4$ are, respectively, given as
 \begin{equation}
  {\bar A}_2
   =
   \frac{2\mycn({\bar b}\frac{\myIm I'}2)\mydn({\bar b}\frac{\myIm I'}2)}
   {1+{\bar k}\mysn({\bar b}\frac{\myIm I'}2)^2},
   ~
   {\bar A}_3
   =
   1,
   ~
   {\bar A}_4
   =
   -\frac{({\bar k}^{\frac12}+{\bar k}^{-\frac12})^2}
   {1+{\bar k}\mysn({\bar b}\frac{\myIm I'}2)^2}.
   \label{eq:4.4}
 \end{equation}
 The ECS in the $Q$-state Potts model is the same as 
 those of the Ising models on the Union Jack and 4-8 lattices 
 \cite{Holzer1990a}.

 \TABLEtwo

 \FIGthree
 
 Note that, for a given $\theta_{\perp}$, $\gamma^{\ast}/\mykB T^{\ast}$
 (or the inverse correlation length $1/\xi$) is a function of $k$ and $Q$.
 From (II), it follows that with the increase of $Q$ but keeping $k$ fixed
 $\gamma^{\ast}/\mykB T^{\ast}$ becomes larger in all directions.
 It is helpful to calculate the radius of curvature $\rho$.
 The row and the diagonal directions are particularly important 
 in connection with the roughening transition phenomena:
 In the zero temperature limit,
 we expect that the ECS is a square,
 and that a facet and a corner appear at $\theta_{\perp}=0$ and $\pi/4$,
 respectively.
 Since
 $\Lambda{\bf R}=\gamma^{\ast}(\cos\theta_{\perp},\sin\theta_{\perp})$
 in these directions,
 it follows that
 \begin{equation}
  \frac{\rho}{|{\bf R}|}
   =
   \left(
    1
    +
    \frac1{\gamma^{\ast}}
    \frac{d^2\gamma^{\ast}}{d{\theta_{\perp}}^2}
   \right),
   \quad
   \theta_{\perp}=0,~\pi/4.
   \label{eq:4.5}
 \end{equation}
 We can proffer the numerical data of
 $\gamma^{\ast}/\mykB T^{\ast}$ and $\rho/|{\bf R}|$
 at
 $\theta_{\perp}=0$ and $\pi/$4. 
 We summarize the results in \myTable{table:2},
 which indicates that, for a given $k$, the ECS deforms slightly to a 
 more circular shape as $Q$ increases.

 The deformation is small: Up to a few percentages in curvature.
 To make it visible, we normalize the bare data using
 the corresponding values of $Q=2$
 \cite{Rottman1981,Avron1982}.
 In \myFig{fig:3},
 the data normalized by the corresponding exact values are plotted.
 \myFigs{fig:3}(a) and \ref{fig:3}(b)
 show that, as $Q$ increases (with $k$ fixed), 
 $\gamma^{\ast}/\mykB T^{\ast}$ becomes larger in both directions.
 From \myFigs{fig:3}(c) and \ref{fig:3}(d),
 we find that the radius of the curvature becomes smaller at
 $\theta_{\perp}=0$, and larger at $\theta_{\perp}=\pi/4$,
 which means that the ECS of the $Q$-state Potts model becomes
 slightly rounded in the facet direction and simultaneously flatter
 in the corner direction.
 Consequently, 
 the ECS deforms monotonically to a more circular shape as
 $Q$ increases
 (see, for example,
 \exFig{4} of \myRef{Fujimoto1996}
 or
 \exFig{3} of \myRef{Holzer1990a}).
 
 The $Q$ dependence of the shape can be extended into $Q\geq1$ with the
 ECS replaced by the polar plot of $1/\xi$.
 The results obtained here is somewhat unusual:
 In typical cases, 
 as the correlation length of the system becomes larger, the ECS or
 the polar plot of $1/\xi$ more circular.
 One should note that the unusual situation also occurs in the
 eight-vertex model
 \cite{Fujimoto1996}
 and the Ising models on the Union Jack and 4-8 lattices
 \cite{Holzer1990}.
 
 In the eight-vertex model, continuously varying exponents can be
 explained by the weak universality concept
 \cite{Suzuki1974},
 where the inverse correlation length $1/\xi$ is regarded as a variable 
 measuring departure from criticality.
 Our results imply that the elliptic
 modulus $k$ describing the ACL is more essential than $1/\xi$.
 That is,
 even if they have different values of $1/\xi$,
 the models sharing the same value of $k$ are the same in the amount
 of the deviation from the critical point.
 We suggest a possibility that the algebraic geometry provides
 the birational equivalence
 \cite{Namba1984}
 as a framework to denote this kind of equivalence.
 Note that the algebraic curve (\ref{eq:1.3}) is a singular curve
 possessing two nodes at infinity, and the algebraic geometry offers a
 standard procedure to treat such curves.
 One scenario is that the weak universality concept is connected with 
 the birational equivalence between algebraic curves like
 \myEq{eq:1.3}.
 It is expected that the connection to the algebraic geometry will break
 a new ground in the study of statistical models.
 
 \subsection{Universal asymptotic forms}
 \label{sec:4.b}

 Before the analyses of the eight-vertex model,
 we commonly observed the curves like \myEq{eq:1.3} as the ECSs of the
 various models solved by the Pfaffian method and that the curves can be
 related to the three properties (i)--(iii);
 see \mySecs{sec:1} and \ref{sec:2}, and references therein.
 In \myRef{Holzer1990a}
 it was also shown that the ECSs like \myEq{eq:1.3}
 do not survive for the modified KDP model because its excitations
 exhibit a unidimensional band structure and explicitly break the double
 periodicity condition.
 These imply that the universality of \myEq{eq:1.3} and the
 applicability of \myEq{eq:3.2} are connected with rather generic
 properties than a specific solvability condition.

 Further, the three properties are expected to be robust against some
 continuous variations of lattice models.
 For the $Q$-state Potts model, by MC simulations, we confirmed that
 \myEq{eq:3.2} can indeed fit the numerical data of asymptotic
 correlation functions with high accuracy.
 This indicates that the universality of our form (\ref{eq:3.2}) emerges
 via the robustness of the three properties (i)--(iii).
 
 Further investigations on this subject are desirable.
 We expect that
 \myEq{eq:3.2} or (\ref{eq:2.3})
 is a universal form for the asymptotic correlation functions with the
 $C_{4v}$ symmetry.
 While the $Q$-state Potts model possesses discrete variables,
 an investigation of continuous spin models, like the
 classical XY model, is important to clarify the degree of applicability
 of \myEq{eq:3.2}.
 In this paper, we have restricted ourselves to the models defined on
 the square lattice.
 It is natural to expect that the same argument is applicable
 to other lattices, e.g., a triangular, a honeycomb, and so on.
 Thus,
 modifications of \myEq{eq:3.2} for other point groups, e.g., $C_{6v}$ are
 interesting
 \cite{Holzer1990a,Vaidya1976,Zia1986,
 Fujimoto1999,Fujimoto1998,Fujimoto2002a};
 see also
 \myRefs{Jayaprakash1984,Holzer1989}.
 At last,
 our investigations may include an application of the present form
 to the problems such as the susceptibility calculations
 containing higher-order terms
 \cite{Yamada1986,Chan2011}.
 We will report our studies on these topics in the future.
 
 \acknowledgements
 We thank Professors
 Macoto Kikuchi and Yutaka Okabe
 for stimulating discussions.
 The main computations were performed using the facilities in
 Tohoku University and Tokyo Metropolitan University.
 This research was supported by a grant-in-aid from KAKENHI
 No.~26400399.
 
 \appendix
 \makeatletter
 \renewcommand{\theequation}{\Alph{section}\arabic{equation}}
 \@addtoreset{equation}{section}
 \makeatother
 
 \section{EXACT CALCULATION OF CORRELATION LENGTH IN SQUARE-LATTICE
 ISING MODEL}
 \label{appendix:A}

 In \exchap{7} of
 \myRef{Baxter1982}
 Baxter exactly calculated the correlation length of the square-lattice
 Ising model along the diagonal direction.
 We extend the transfer matrix argument into all directions using the
 shift operator.
 In order to find the role of the $C_{4v}$ symmetry, we consider the
 Ising model on a rotated square lattice.
  
 \subsection{Commuting transfer matrices in $\boldsymbol{Q=2}$}
 \label{appendix:A1}
 
 Suppose a square lattice drawn diagonally.
 For each site ${\bf r}=i{\bf e}_x+j{\bf e}_y$ we introduce a variable
 $\sigma_{\bf r}$,
 which is related to the $Q$-valued variable $q_{\bf r}$ in
 \myEq{eq:1.1}
 as $\sigma_{\bf r}=\exp(2\pi\myIm q_{\bf r}/Q)$.
 When $Q=2$,
 $\sigma_{\bf r}=\pm 1$
 and
 $\delta\left(q_{\bf r},q_{{\bf r}'}\right)=
 \frac12(1+\sigma_{\bf r}\sigma_{{\bf r}'})$.
 Thus,
 the $Q=2$ Potts model
 (with $J_{{\bf r, r}'}$ replaced by $2J_{{\bf r, r}'}$)
 is equivalent to the Ising model whose Hamiltonian is given by
 \begin{equation}
  E
   =
   -\sum_{\bf r}
   \left(
    J \sigma_{{\bf r}+{\bf e}_x}\sigma_{\bf r}+
    J'\sigma_{{\bf r}+{\bf e}_y}\sigma_{\bf r}
   \right).
   \label{eq:A1}
 \end{equation}
 The nearest-neighbor spins are coupled by $J$ or $J'$ depending on
 the direction.
 We assume $J$, $J'>0$. 
 The partition function is 
 \begin{equation}
  Z
   =
   \sum_{\sigma}
   \exp\left[
	\sum_{\bf r}
	\left(
	 K \sigma_{{\bf r}+{\bf e}_x}\sigma_{\bf r}+
	 K'\sigma_{{\bf r}+{\bf e}_y}\sigma_{\bf r}
	\right)
       \right],
   \label{eq:A2}
 \end{equation}
 where the outer sum is over all spin configurations and
 the reduced couplings
 $K =J /\mykB T$ and
 $K'=J'/\mykB T$.

 We introduce diagonal-to-diagonal transfer matrices:
 Consider two successive rows, and let
 $\sigma =\{\sigma_0 ,\dots,\sigma_{N-1} \}$ (respectively,
 $\sigma'=\{\sigma_0',\dots,\sigma_{N-1}'\}$)
 be the spins in the lower (respectively, upper) row.
 We assume the periodic boundary conditions in both directions.
 Then, as shown in \myFig{fig:A1}(a),
 the transfer matrices ${\bf V}$ and ${\bf W}$ are defined by
 elements as
 \begin{align}
  [{\bf V}]_{\sigma,\sigma'}
  &= 
  \exp
  \left[
  \sum_{l=0}^{N-1}(K\sigma_{l+1}\sigma_l'+K'\sigma_l\sigma_l')
  \right],
  \cr
  [{\bf W}]_{\sigma,\sigma'}
  &= 
  \exp
  \left[
  \sum_{l=0}^{N-1}(K\sigma_l\sigma_l'+K'\sigma_l\sigma_{l+1}')
  \right],
  \label{eq:A3}
 \end{align}
 where
 $\sigma_N=\sigma_0$ and $\sigma_N'=\sigma_0'$
 (see \exchap{7} of \myRef{Baxter1982}).
 When the system has $2M$ rows,
 the partition function is given as follows:
 \begin{equation}
  Z
   =
   \myTr{\bf U}^{M}
   =
   \sum_{p=0}^{2^N-1}\left(\Lambda_p^2\right)^M,
   \qquad
   {\bf U}={\bf VW},
   \label{eq:A4}
 \end{equation}
 where $\Lambda_p^2$ is the $p$th eigenvalue of ${\bf U}$.
  
 \FIGAone
  
 Above the critical temperature $\myTc$,
 we parametrize $K$ and $K'$ using the elliptic functions with the
 modulus $k\in(0,1)$ as
 \begin{align}
  &\sinh2K
  =
  \frac{k\mysn(\myIm u)}{\myIm},
  \quad
  \cosh2K
  =
  \mydn(\myIm u),
  \cr
  &\sinh2K'
  =
  \frac{\myIm}{\mysn(\myIm u)},
  \quad
  \cosh2K'
  =
  \myIm\frac{\mycn(\myIm u)}{\mysn(\myIm u)}.
  \label{eq:A5}
 \end{align}
 The quarter periods are denoted by $I$ and $I'$;
 and the argument $u$ satisfies the condition $0<u<I'$
 (see also \exchap{15} of \myRef{Baxter1982} and \myRef{Baxter1978}).
 For $T<\myTc$,
 we find the similar parametrization:
 \begin{align}
  &\sinh2K
  =
  \frac{\mysn(\myIm u)}{\myIm},
  \quad
  \cosh2K
  =
  \mycn(\myIm u),
  \cr
  &\sinh2K'
  =
  \frac{\myIm}{k\mysn(\myIm u)},
  \quad
  \cosh2K'
  =
  \myIm\frac{\mydn(\myIm u)}{k\mysn(\myIm u)}.
  \label{eq:A6}
 \end{align}

 Regard $k$ as a fixed constant, and $u$ as a complex variable.
 Then
 ${\bf U}$ is a function of $u$.
 When we write it as ${\bf U}(u)$, it satisfies the commutation
 relation
 \begin{equation}
  [{\bf U}(u),{\bf U}(u')]=0
   \quad
   \forall u, u' \in\mathbb{C}.
   \label{eq:A7}
 \end{equation}
 Further,
 it commutes with the matrix ${\bf R}$ defined by elements as
 \begin{equation}
  [{\bf R}]_{\sigma,\sigma'}
   =
   \prod_{l=0}^{N-1}
   \delta(\sigma_l,-\sigma_l'),
   \label{eq:A8}
 \end{equation}
 i.e.,
 \begin{equation}
  [{\bf U}(u),{\bf R}]=0.
   \label{eq:A9}
 \end{equation}

 For simplicity,
 suppose that $N$ is an even number,
 then it follows that $\Lambda(u)$ is a doubly periodic function of
 $u$:
 \begin{align}
  &
  \Lambda(u+2I')=r\Lambda(u),~
  \Lambda(u-2\myIm I)=r\Lambda(u)
  \text{~for~}T>\myTc,
  \label{eq:A10}
  \\
  &
  \Lambda(u+2I')=r\Lambda(u),~
  \Lambda(u-2\myIm I)=\Lambda(u)
  \text{~for~}T<\myTc,
  \label{eq:A11}
 \end{align}
 where $r$ ($=\pm1$) is the eigenvalue of ${\bf R}$ corresponding to
 $\Lambda(u)$.
 In addition, it is found that
 \begin{align}
  &\Lambda(u)\Lambda(u+I')
  \cr
  &=
  (-2)^N
  \Bigl\{\frac1{\mysn(\myIm u)^{N}}+\left[k\mysn(\myIm u)\right]^N r\Bigr\}
  \text{~for~}T>\myTc,
  \label{eq:A12}
  \\
  &=
  (-2)^N
  \Bigl\{\frac1{[k\mysn(\myIm u)]^{N}}+\mysn(\myIm u)^N r\Bigr\}
  \text{~for~}T<\myTc.
  \label{eq:A13}
 \end{align}

 To determine explicit forms of $\Lambda(u)$,
 we can use
 \myEq{eq:A12} with the periodicity (\ref{eq:A10}) for $T>\myTc$,
 and
 \myEq{eq:A13} with \myEq{eq:A11}  for $T<\myTc$.
 For example,
 it is shown that the maximum eigenvalue
 $\Lambda_0(u)^2$ behaves as
 $\Lambda_0(u)^2\sim\kappa(u)^{2N}$, when $N$ becomes large,
 and the per-site free energy $f$ is given by
 \begin{align}
  -&\frac{f}{\mykB T}
  =
  \ln\kappa(u)
  =
  \cr
  &
  \frac{1}{2\pi}  
  \int_0^\pi d\theta 
  \ln\left\{2\left[\cosh2K\cosh2K'+c(\theta)\right]\right\},
  \quad
  \label{eq:A14}
 \end{align}
 where
 \begin{align}
  c(\theta)
  &=
  \left(1-2k^{+1}\cos2\theta+k^{+2}\right)^{\frac12}
  \text{~~for~~}T>\myTc,
  \label{eq:A15}
  \\
  &=
  \left(1-2k^{-1}\cos2\theta+k^{-2}\right)^{\frac12}
  \text{~~for~~}T<\myTc
  \label{eq:A16}
 \end{align}
 (see \exSec{7.9} of \myRef{Baxter1982}).

 \subsection{Shift operator method}
 \label{appendix:A2}

 In \exchap{7} of \myRef{Baxter1982}
 Baxter analyzed the asymptotic behavior of the 
 correlation function using ${\bf U}(u)$.
 The correlation length was exactly calculated along the diagonal
 direction.
 We can extend the calculation into all directions with the help of the
 shift operator ${\bf S}$, which moves spins on a row along horizontal
 direction, i.e.,
 \begin{equation}
  [{\bf S}]_{\sigma,\sigma'}
   =
   \prod_{l=0}^{N-1}
   \delta(\sigma_l,\sigma_{l+1}').
   \label{eq:A17}
 \end{equation}
 
 Let ${\bf r^+}$ be the position vector of a site on the sublattice
 containing the origin ${\bf o}$; 
 we start with choosing two sites on the same sublattice,
 but this restriction will be removed later.
 In the usual transfer matrix method the expectation value of the spin
 product $\sigma_{\bf o}\sigma_{{\bf r}^+}$ is represented as
 \begin{align}
  &\myav{\sigma_{\bf o}\sigma_{{\bf r}^+}}
  =
  \frac{\myTr[{\bf A}_0{\bf U}^m{\bf A}_n{\bf U}^{M-m}]}
  {\myTr{\bf U}^M},
  \cr
  &
  ~~
  {{\bf r}^+}
  =
  n({\bf e}_x+{\bf e}_y)+m(-{\bf e}_x+{\bf e}_y)
  \label{eq:A18}
 \end{align}
 where
 ${\bf A}_k$s are defined by
 \begin{equation}
  [{\bf A}_k]_{\sigma,\sigma'}
   =
   \sigma_k
   \prod_{l=0}^{N-1}
   \delta(\sigma_l,\sigma_l').
   \label{eq:A19}
 \end{equation}

 Apply a similarity transformation to diagonalize ${\bf U}$.
 We take the $M\rightarrow\infty$ limit first, then the
 $N\rightarrow\infty$ limit.
 In the $M\rightarrow\infty$ limit, we find that
 \begin{equation}
  \myav{\sigma_{\bf o}\sigma_{{\bf r}^+}}
   =
   \sum_p
   [{\tilde {\bf A}}_0]_{0,p}
   [{\tilde {\bf A}}_n]_{p,0}
   \left[\frac{\Lambda_p(u)}{\Lambda_0(u)}\right]^{2m},
   \label{eq:A20}
 \end{equation}
 where $\Lambda_p(u)^2$ is the $p$th eigenvalue of ${\bf U}(u)$ in
 decreasing order of magnitude, and ${\tilde{\bf A}}_k$ is the matrix
 transformed from ${\bf A}_k$.
 \myEquation{eq:A20} implies that the ratios between the eigenvalues of
 ${\bf U}(u)$ essentially determine the asymptotic behavior of the
 correlation function along the diagonal direction.
 For example,
 when $n$ is fixed and $m$ becomes large, the correlation length along
 the diagonal direction is calculated from the ratios between
 $\Lambda_0(u)^2$ and the next-largest eigenvalues.

 To find the asymptotic form in all directions, we consider the 
 anisotropic correlation length (ACL), 
 which is obtainable by taking the $m\rightarrow\infty$ limit with the
 ratio $n/m$ fixed to be constant.
 In this limit contribution from the matrix elements
 $[{\tilde {\bf A}}_0]_{0,p}$ and
 $[{\tilde {\bf A}}_n]_{p,0}$
 is important as well as the ratios between the eigenvalues.
 This causes a difficulty since the direct calculation of the matrix
 elements is very complicated in most cases.

 We can overcome the difficulty with the help of the shift operator
 ${\bf S}$
 \cite{Fujimoto1990,Fujimoto1990a,Kluemper1990};
 see also
 \myRefs{Yamada1983,Yamada1984}. 
 Because the shift operator relates ${\bf A}_n$ to ${\bf A}_0$ as 
 \begin{equation}
  {\bf A}_n={\bf S}^{-n}{\bf A}_0{\bf S}^n,
   \label{eq:A21}
 \end{equation}
 we rewrite \myEq{eq:A20} as
 \begin{equation}
  \myav{\sigma_{\bf o}\sigma_{{\bf r}^+}}
   =
   \sum_p
   [{\tilde{\bf A}}_0]_{0,p}
   [{\tilde{\bf A}}_0]_{p,0}
   \left[
    \frac{\Lambda_p(u)}{\Lambda_0(u)}
    \left({\frac{S_p}{S_0}}\right)^{\frac\gamma2}
   \right]^{2m}
   \label{eq:A22}
 \end{equation}
 with
 \begin{equation}
  \gamma=-\frac{n}{m},
   \label{eq:A23}
 \end{equation}
 where $S_p$ is the $p$th eigenvalue of ${\bf S}$ and $S_0=1$.
 \myEq{eq:A22} shows that we can obtain the ACL from the
 eigenvalues of ${\bf U}(u)$ and those of ${\bf S}$ without
 calculating the matrix elements.

 \subsection{Limiting function $\boldsymbol{L(u)}$}
 \label{appendix:A3}
 
 To consider the $N\rightarrow\infty$ limit, we define limiting
 functions as
 \begin{equation}
  L_p(u)
   =
   \lim_{N\rightarrow\infty}\frac{\Lambda_p(u)}{\kappa(u)^N},
   \quad p=0,1,\dots,
   \label{eq:A24}   
 \end{equation} 
 where $\kappa(u)$ is given by \myEqs{eq:A14}--(\ref{eq:A16});
 note that $L_0(u)\equiv1$.
 It is shown that
 \begin{equation}
  \frac{S_p}{S_0}=\lim_{u\rightarrow 0}L_p(u)^2.
   \label{eq:A25}
 \end{equation}
 From \myEqs{eq:A22}, (\ref{eq:A24}), and (\ref{eq:A25}),
 we obtain
 \begin{equation}
  \myav{\sigma_{\bf o}\sigma_{{\bf r}^+}}
   =
   \sum_p
   [{\tilde {\bf A}}_0]_{0,p}
   [{\tilde {\bf A}}_0]_{p,0}
   \left[L_p(u)L_p(0)^{\gamma}\right]^{2m}.
   \label{eq:A26}
 \end{equation}
 
 Using \myEqs{eq:A10}--(\ref{eq:A16}),
 we can determine the form of $L(u)$.
 When $N$ becomes large and $-I'/2<\Re(u)<I'/2$,
 the first term is dominant on the right-hand side of
 \myEq{eq:A12}.
 We keep only the dominant term there
 \cite{Kluemper1990}.
 Divide the both sides by $\kappa(u)^N\kappa(u+I')^N$ and use
 \myEqs{eq:A14}, (\ref{eq:A15}), and (\ref{eq:A24}).
 Combining the result from the second equation of (\ref{eq:A10}),
 we find that
 \begin{equation}
  L(u)L(u+I')=1,~
   L(u-2\myIm I)=rL(u)
   \text{~for~}T>\myTc.
   \label{eq:A27}
 \end{equation}
 Similarly,
 from \myEq{eq:A13} and the second equation of (\ref{eq:A11}),
 we obtain
 \begin{equation}
  L(u)L(u+I')=1,~
   L(u-2\myIm I)=L(u)
   \text{~for~}T<\myTc.
   \label{eq:A28}
 \end{equation}
 
 Because the zeros of $\Lambda_0(u)$ are located on the line
 $\Re(u)=-I'/2$ in a periodic rectangle, the first equation
 of (\ref{eq:A27}) or (\ref{eq:A28}) shows that the limiting function
 is written as
 \begin{equation}
  L(u)
   =
   F(u)
   \prod_{l=1}^\nu
   \frac{a-xa_l}{a-x^3a_l},~~
   -\frac{I'}{2}<\Re(u)\leq\frac{3I'}{2}
   \label{eq:A29}
 \end{equation}
 with
 $a=\exp\left(-{\pi u}/{I}\right)$,
 $a_l=\exp\left({\myIm\pi\phi_l}/{I}\right)$,
 and
 $x=\exp\left(-{\pi I'}/{2I}\right)$.
 A function $F(u)$ is analytic and nonzero for
 $-I'/2<\Re(u)\leq3I'/2$.
 Thus,
 the limiting functions are labeled by an integer $\nu$ and real
 numbers $\phi_1,\phi_2,\dots,\phi_{\nu}$ instead of $p$.
 
 Substitute
 \myEq{eq:A29}
 into the first equation of
 (\ref{eq:A27}) or (\ref{eq:A28}),
 take the logarithms of both sides, and then
 expand them in the annulus $-I'/2<\Re(u)<I'/2$ using the form
 \begin{align}
  \ln F(u)
  =
  c_0
  +
  &\alpha\ln a
  +
  \sum_{\mu=1}^\infty\left(c_{\mu} a^{\mu}+c_{-\mu}a^{-\mu}\right),
  \cr
  &-\frac{I'}{2}<\Re(u)\leq\frac{3I'}{2}.
  \label{eq:A30}
 \end{align}
 Equating coefficients gives
 \begin{align}
  &
  c_0
  =
  -\frac12\sum_{l=1}^{\nu}\ln(-a_l),
  ~
  \alpha
  =
  \frac\nu2,
  ~
  c_{\mu}
  =
  \frac{x^{\mu}}{1+x^{2\mu}}\frac{1}{\mu}\sum_{l=1}^{\nu}a_l^{-\mu},
  \cr
  &
  c_{-\mu}
  =
  -\frac{x^{3\mu}}{1+x^{-2\mu}}\frac{1}{\mu}\sum_{l=1}^{\nu}a_l^{\mu}
  \label{eq:A31}
 \end{align}
 with $\mu=1,2,\dots$.
 We find that
 \begin{equation}
  L(u)
   =
   \pm\prod_{l=1}^{\nu}k^\frac12\mysn(\myIm u-\phi_l-\frac{\myIm I'}{2}),
   ~~
   -\frac{I'}{2}<\Re(u)\leq\frac{3I'}{2}.
   \label{eq:A32}
 \end{equation}

 For $T>\myTc$,
 from the second equation of (\ref{eq:A27}), it follows that
 $\nu$ is an odd (even) integer if $r=-1$ ($r=+1$).
 The next-largest eigenvalues correspond to the case with $\nu=1$ and
 $r=-1$.
 
 For $T<\myTc$,
 the two largest eigenvalues $\Lambda_0(u)^2$ and $\Lambda_1(u)^2$ are
 asymptotically degenerate when $N$ becomes large.
 Note that
 $r=+1$ for $\Lambda_0(u)^2$ and
 $r=-1$ for $\Lambda_1(u)^2$
 (see \exSec{7.10} of \myRef{Baxter1982}).
 The second equation of (\ref{eq:A28}) shows that $\nu$ is an even
 number.
 We thus find that the next-largest eigenvalues correspond to
 the cases with $\nu=2$ and $r=\pm1$.

 \subsection{Anisotropic correlation lengths $\boldsymbol{Q=2}$}
 \label{appendix:A4}
 
 It is shown that, because of continuous distributions of
 eigenvalues, the sum in \myEq{eq:A22}
 becomes integrals over
 $\phi_l$s in the $N\rightarrow\infty$ limit
 \cite{Johnson1973}.
 For simplicity, 
 we choose the positive sign in \myEq{eq:A32}.
 Detailed analysis also shows that the maximum eigenvalue
 $\Lambda_0(u)^2$ corresponds to the case 
 $r_0=+1$,
 and
 $[{\tilde {\bf A}}_0]_{0,p}$
 and
 $[{\tilde {\bf A}}_0]_{p,0}$
 vanish unless
 $r_p=-1$
 due to the ${\mathbb Z}_2$ symmetry of the system,
 where $r_p$ is the $p$th eigenvalue of ${\bf R}$. 

 For $T>\myTc$,
 only the band of next-largest eigenvalues with $\nu=1$ and $r=-1$
 contributes to the leading asymptotic behavior of the correlation
 function in the limit of $m$ large with $\gamma$ fixed.
 It follows that
 \begin{widetext}
 \begin{equation}
  \myav{\sigma_{\bf o}\sigma_{{\bf r}^+}}-
   \myav{\sigma_{\bf o}}\myav{\sigma_{{\bf r}^+}}
   \sim
   \int^{I}_{-I}d\phi\rho(\phi)
   \left\{k^\frac12\mysn(\myIm u-\phi-\frac{\myIm I'}{2})
    \left[k^\frac12\mysn(-\phi-\frac{\myIm I'}{2})\right]^\gamma
   \right\}^{2m},
   \label{eq:A33}
 \end{equation}
 \end{widetext}
 where the function $\rho(\phi)$ is to be determined from the
 distribution of eigenvalues, and matrix elements
 $[{\tilde {\bf A}}_0]_{0,p}$,
 $[{\tilde {\bf A}}_0]_{p,0}$.
 Because $r=+1$ for eigenvalues with $\nu=2$, 
 $[{\tilde {\bf A}}_0]_{0,p}$
 and
 $[{\tilde {\bf A}}_0]_{p,0}$ vanish.
 Therefore,
 the first correction to the asymptotic behavior (\ref{eq:A33}) comes
 from the integral over the band of eigenvalues with $\nu=3$ and $r=-1$
 \cite{Yamada1984}.
 
 As stated in \myAppendix{appendix:A2},
 we extend the above analysis to include any pair of sites. 
 Because ${\bf r}^+=(n-m){\bf e}_x+(n+m){\bf e}_y$
 [see \myFig{fig:A1}(b)], 
 we obtain the transformation of the coordinates, i.e.,
 \begin{equation}
  i=n-m,
   \qquad
   j=n+m.
   \label{eq:A34}
 \end{equation}
 We can remove the restriction $i\pm j={\rm even}$ in \myEq{eq:A33}
 to find the correlation function for all ${\bf r}$ as
 \begin{widetext}
 \begin{equation}
  \myav{\sigma_{\bf o}\sigma_{\bf r}}-
   \myav{\sigma_{\bf o}}\myav{\sigma_{\bf r}}
   \sim
   \int^I_{-I}d\phi\rho(\phi)
   \left\{
    k\mysn(\myIm u-\phi-\frac{\myIm I'}{2})\mysn(-\phi+\frac{\myIm I'}{2})
    \left[
     k\mysn(\myIm u-\phi+\frac{\myIm I'}{2})\mysn(-\phi+\frac{\myIm I'}{2})
    \right]^{\Gamma}
   \right\}^j,
   \label{eq:A35}
 \end{equation}
 \end{widetext}
 where $\Gamma$ is the ratio given by
 \begin{equation}
  \Gamma=\frac{i}{j}=\frac{\gamma+1}{\gamma-1}.
   \label{eq:A36}
 \end{equation}

 Along the direction designated by $\Gamma$, the correlation length
 $\xi$ is defined as
 \begin{equation}
  -\frac{1}{\xi}
   =
   \lim_{R\rightarrow\infty}
   \frac{\ln
   \left[
    \myav{\sigma_{\bf o}\sigma_{\bf r}}-
    \myav{\sigma_{\bf o}}\myav{\sigma_{\bf r}}
   \right]}{R},
   \quad 
   R=\sqrt{i^2+j^2},
   \label{eq:A37}
 \end{equation}
 where the limit is taken with $\Gamma$ fixed.
 We regard $\xi$ as a function of $\theta$, the angle between
 ${\bf e}_x$ and the direction of $\Gamma$.
 Explicitly, $\Gamma$ is related to $\theta$ as
 \begin{equation}
  \Gamma=\frac1{\tan\theta},
   \quad
   \frac{\pi}{4}<\theta<\frac{5\pi}{4}. 
   \label{eq:A38}
 \end{equation}
 We assume an analyticity of $\rho(\phi)$ and then estimate the
 integral on the right-hand side of \myEq{eq:A35} by the method of
 steepest descent.
 It follows that
 \begin{align}
  -\frac1\xi
  &=
  \sin\theta
  \ln
  \left[
  k
  \mysn(\myIm u-\phi_{\rm s}-\frac{\myIm I'}{2})
  \mysn(-\phi_{\rm s}+\frac{\myIm I'}{2})
  \right]
  \cr
  &+ 
  \cos\theta
  \ln
  \left[
  k
  \mysn(\myIm u-\phi_{\rm s}+\frac{\myIm I'}{2})
  \mysn(-\phi_{\rm s}+\frac{\myIm I'}{2})
  \right],
  \label{eq:A39}
 \end{align}
 where the saddle point $\phi_{\rm s}$ is determined as a function of
 $\theta$ by
 \begin{align}
  &\sin\theta\frac{d}{d\phi_{\rm s}}
  \ln
  \left[
  k
  \mysn(\myIm u-\phi_{\rm s}-\frac{\myIm I'}{2})
  \mysn(-\phi_{\rm s}+\frac{\myIm I'}{2})
  \right]
  +
  \cr
  &\cos\theta\frac{d}{d\phi_{\rm s}}
  \ln
  \left[
  k
  \mysn(\myIm u-\phi_{\rm s}+\frac{\myIm I'}{2})
  \mysn(-\phi_{\rm s}+\frac{\myIm I'}{2})
  \right]
  =
  0
  \label{eq:A40}
 \end{align}
 with the condition 
 $\phi_{\rm s}=\myIm u-{\myIm I'}/{2}\pm I$ for
 $\theta={3\pi}/{4}$.
 The relation $\xi(\theta+\pi)=\xi(\theta)$ implies
 that the result in \myEqs{eq:A39} and (\ref{eq:A40})
 is analytically continued into $0<\theta<2\pi$.
 Note that increasing $\theta$ by $2\pi$ causes $\Im(\phi_{\rm s})$
 to decrease by $2I'$.
 We expect that $\rho(\phi)$ is a doubly periodic function and
 is analytic inside and on a periodic rectangle.
 According to Liouville's theorem, it should be a constant.

 Shifting the integration path along the imaginary axis, 
 we can rewrite \myEq{eq:A35} as
 \begin{align}
  &
  \myav{\sigma_{\bf o}\sigma_{\bf r}}-
  \myav{\sigma_{\bf o}}
  \myav{\sigma_{\bf r}}
  \sim
  \text{Const}
  \cr
  &
  \times\oint\frac{d\alpha}{\alpha}
  \oint\frac{d\beta}{\beta}
  \frac{\alpha^i \beta^j}
  {2a-\gamma_1(\alpha+\alpha^{-1})-\gamma_2(\beta+\beta^{-1})},\qquad
  \label{eq:A41}
 \end{align}
 where contours of integrations are unit circles, and
 \begin{equation}
  a=(1+z_1^2)(1+z_2^2),
   \quad
   \gamma_1=2z_2(1-z_1^2),
   \quad
   \gamma_2=2z_1(1-z_2^2)
   \label{eq:A42}
 \end{equation}
 with
 $z_1=\tanh K$ and $z_2=\tanh K'$.
 We note that
 \myEqs{eq:A41} and (\ref{eq:A42}) coincide with the results in
 \exSec{4} of \myRef{Cheng1967}
 and
 \exSec{XII-4} of \myRef{McCoy1973};
 see also
 \myRef{Yamada1984}.
 In the case of the isotropic interactions the denominator of the
 integrand has the same form as that in a special case of the left-hand
 side of \myEq{eq:1.3}.
 Therefore,
 it follows that the structure factor of the asymptotic correlation
 function possesses the same algebraic property as that of the
 eight-vertex model.

 For $T<\myTc$,
 because
 the band of next-largest eigenvalues with $\nu=2$ and $r=-1$
 determines the leading asymptotic behavior of the correlation
 function.
 We obtain
\begin{widetext}
 \begin{align}
  \myav{\sigma_{\bf o}\sigma_{{\bf r}^+}}-
  \myav{\sigma_{\bf o}}
  \myav{\sigma_{{\bf r}^+}}
  \sim
  \int^{I}_{-I}d\phi_1\int^{I}_{-I}d\phi_2\rho(\phi_1,\phi_2)
  &
  \left\{
  k^\frac12\mysn(\myIm u-\phi_1-\frac{\myIm I'}{2})
  \left[k^\frac12\mysn(-\phi_1-\frac{\myIm I'}{2})\right]^{\gamma}
  \right\}^{2m}
  \cr
  \times
  &
  \left\{
  k^\frac12\mysn(\myIm u-\phi_2-\frac{\myIm I'}{2})
  \left[k^\frac12\mysn(-\phi_2-\frac{\myIm I'}{2})\right]^{\gamma}
  \right\}^{2m}.
  \label{eq:A43}
 \end{align}
\end{widetext}
 Again,
 the function $\rho(\phi_1,\phi_2)$ is to be calculated from the
 distribution of eigenvalues and the matrix elements
 $[{\tilde {\bf A}}_0]_{0,p}$,
 $[{\tilde {\bf A}}_0]_{p,0}$. 
 Assume an analyticity of $\rho(\phi_1,\phi_2)$ and integrate by
 steepest descents.
 Then,
 we find that the correlation length $\xi^{\ast}$ below $\myTc$
 is related to $\xi$ above $\myTc$ determined by
 \myEqs{eq:A39} and (\ref{eq:A40}) as
 \begin{equation}
  \xi=2\xi^{\ast}
   \quad
   \text{in all directions}.
   \label{eq:A44}
 \end{equation}

 Shifting the integration paths suitably, we find that \myEq{eq:A43}
 is essentially the same as the leading asymptotic form in
 \exSec{3} of \myRef{Cheng1967}
 and
 \exSec{XII-3} of \myRef{McCoy1973}.
 The asymptotic correlation function is expressed in terms of the
 differential forms on the same algebraic curve as in \myEq{eq:A41}.
 The difference from the case $T>\myTc$ is that two elliptic curves are
 needed in the case $T<\myTc$.
 
 \subsection{Passive rotations}
 \label{appendix:A5}

 In
 \myRef{Yamada1984}
 it was shown that
 the results of the correlation functions by the Pfaffian method in
 \myRefs{McCoy1973,Cheng1967}
 are equivalent to those by the row-to-row transfer matrix.
 The analyses in the previous section suggest that
 difference in direction along which the transfer matrix is defined
 causes a shift or deformation of the integration paths in
 the asymptotic correlation function.
 To clarify this point,
 we apply the argument for the eight-vertex model in
 \myRef{Fujimoto2002}
 to the square-lattice Ising model.
 
 The method given in \myAppendix{appendix:A2} corresponds to the active
 rotations.
 We employ another method corresponding to the passive rotations: 
 We define the Ising model on a square lattice rotated through an
 arbitrary angle with respect to the coordinate axes.
 The rotated system is related to an inhomogeneous system possessing a
 one-parameter family of commuting transfer matrices.
 A product of commuting transfer matrices can be interpreted as a
 transfer matrix acting on zigzag walls in the rotated system
 \cite{Fujimoto2002,Fujimoto1994}.

 For convenience, we denote the Boltzmann weight of four edges as
 \begin{equation}
  W(a,b|c,d|u)
   =
   2\cosh(K'a+Kb+K'c+Kd),
   \label{eq:A45}
 \end{equation}
 where $a$, $b$, $c$, and $d$ are the nearest-neighbor spins of $f$
 arranged as in \myFig{fig:A2}.
 Note that $K$, $K'$ are given
 by \myEq{eq:A5} for $T>\myTc$
 and
 by \myEq{eq:A6} for $T<\myTc$.
 
 \FIGAtwo

 The weight $W(a,b|c,d|u)$ satisfies the following properties
 \cite{Fujimoto1994}, i.e.,
 the standard initial condition
 \begin{equation}
  \lim_{u\rightarrow 0}\frac{W(a,b|c,d|u)}{\kappa(u)^2}=\delta(a,c),
   \label{eq:A46}
 \end{equation}
 and the crossing symmetry
 \begin{equation}
  W(a,b|c,d|I'-u)=W(b,a|d,c|u),
   \label{eq:A47}
 \end{equation}
 where $\kappa(u)$ is given by \myEqs{eq:A14}--(\ref{eq:A16}).
 Since $\kappa(I'-u)=\kappa(u)$,
 it follows from
 \myEqs{eq:A46} and (\ref{eq:A47})
 that
 \begin{equation}
  \lim_{u\rightarrow I'}\frac{W(a,b|c,d|u)}{\kappa(u)^2}=\delta(b,d).
   \label{eq:A48}
 \end{equation}
 
 To calculate $\xi$ along the direction $\theta$,
 we consider the Ising model on a square lattice rotated through
 $3\pi/4-\theta$ with respect to the one drawn diagonally.
 Let
 $\sigma =\{\sigma_0,\dots,\sigma_{N-1} \}$
 (respectively,
 $\sigma'=\{\sigma_0',\dots,\sigma_{N-1}'\}$)
 be the spins on the lower (respectively, upper) row of open circles
 shown in \myFig{fig:A1}(a).
 Suppose that $N=(|n|+m)N_0$, where $m>0$ and $N_0$ is an even number.
 Then, 
 we define inhomogeneous transfer matrices as
\begin{widetext}
 \begin{equation}
  [{\bf U}_{\rm IH}(u)]_{\sigma,\sigma'}
   =
   \prod_{l=0}^{N_0-1}
   \left[
    \prod_{s=l(|n|+m)}^{l(|n|+m)+|n|-1}\!\!\!\!\!\!
    W(\sigma_s,\sigma_{s+1}|\sigma_{s+1}',\sigma_s'|u)\!\!\!
    \prod_{t=l(|n|+m)+|n|}^{(l+1)(|n|+m)-1}\!\!\!\!\!\!
    W(\sigma_{t},\sigma_{t+1}|\sigma_{t+1}',\sigma_{t}'|u+u_0-H(-n)I')
   \right]
   \label{eq:A49}
 \end{equation}
 with
 $\sigma_{N} =\sigma_0 $ and
 $\sigma_{N}'=\sigma_0'$,
 where $0<u_0<I'$ and $H(\cdot)$
 is the Heaviside step function.
 
 The commutation relation (\ref{eq:A7}) is generalized as
 \begin{equation}
  [{\bf U}_{\rm IH}(u),{\bf U}_{\rm IH}(u')]=0
   \quad
   \forall u, u' \in\mathbb{C},
   \label{eq:A50}
 \end{equation}
 and (\ref{eq:A9}) as
 \begin{equation}
  [{\bf U}_{\rm IH}(u), {\bf R}]=0.
   \label{eq:A51}
 \end{equation}

 By using ${\bf U}_{\rm IH}(u)$,
 we can construct a transfer matrix ${\bar {\bf U}}$ acting on zigzag
 walls in the rotated system as
 \begin{align}
  \bar{\bf U}
  =
  \left[
  \lim_{u\rightarrow 0}\frac{{\bf U}_{\rm IH}(u)}{\kappa(u)^{2|n|N_0}}
  \right]^m
  \left[
  \lim_{u\rightarrow I'-u_0}\frac{{\bf U}_{\rm IH}(u)}{\kappa(u+u_0)^{2mN_0}}
  \right]^{|n|}\qquad
  &\text{~for~}n>0,
  \cr
  =
  \left[
  \lim_{u\rightarrow I'}\frac{{\bf U}_{\rm IH}(u)}{\kappa(u)^{2|n|N_0}}
  \right]^m
  \left[
  \lim_{u\rightarrow I'-u_0}\frac{{\bf U}_{\rm IH}(u)}{\kappa(u+u_0-I')^{2mN_0}}
  \right]^{|n|}
  &\text{~for~}n\le0,
  \label{eq:A52}
 \end{align}
 where $n$ and $m$ are related to $\theta$ by
 $n/m=\tan(3\pi/4-\theta)$ with $\pi/4<\theta<5\pi/4$
 (see \exFig{2} of \myRef{Fujimoto1994}).
 ${\bar {\bf U}}$ reduces
 to the diagonal-to-diagonal transfer matrix in the case $n=0$, and
 to the row-to-row transfer matrix in the case $n=m$ (or $n=-m$).
 We can find the correlation length along any direction of $\theta$
 from the eigenvalues of ${\bar {\bf U}}$.

 Noting the relations 
 \begin{align}
  &\sum_f W(a,b|f,d|u)W(f,b|c,d|-u)=-(2\sinh2K')^2\delta(a,c),
  \cr
  &\sum_f W(a,b|f,d|u-I')W(f,b|c,d|-u+I')=-(2\sinh2K)^2\delta(a,c)
  \label{eq:A53}
 \end{align}
 with $\sinh2K$, $\sinh2K'$ given by \myEq{eq:A5} or (\ref{eq:A6}),
 we also construct a shift operator ${\bar {\bf S}}$ as
 \begin{align}
  \bar{\bf S}
  =
  \left[-\lim_{u\rightarrow -u_0}(2\sinh2K')^2\right]^{-|n|mN_0}
  \left[
  \lim_{u\rightarrow -u_0}
  \frac{{\bf U}_{\rm IH}(u)}{\kappa(u+u_0)^{2mN_0}}
  \right]^m
  \left[
  \lim_{u\rightarrow 0}\frac{{\bf U}_{\rm IH}(u)}{\kappa(u)^{2|n|N_0}}
  \right]^{|n|}\qquad~
  &\text{~for~}n>0,
  \cr
  =
  \left[-\lim_{u\rightarrow -u_0}(2\sinh2K)^2\right]^{-|n|mN_0}
  \left[
  \lim_{u\rightarrow I'-u_0}
  \frac{{\bf U}_{\rm IH}(u)}{\kappa(u+u_0-I')^{2mN_0}}
  \right]^m
  \left[
  \lim_{u\rightarrow 0}\frac{{\bf U}_{\rm IH}(u)}{\kappa(u)^{2|n|N_0}}
  \right]^{|n|}
  &\text{~for~}n\le0.
  \label{eq:A54}
 \end{align}
\end{widetext}

 We denote eigenvalues of
 ${\bf U}_{\rm IH}(u)$ as $\Lambda_{\rm IH}(u)^2$.
 When $N_0$ (or $N$) becomes large with $n$ and $m$ fixed, the
 maximum eigenvalue $\Lambda_{\rm IH;0}(u)^2$ behaves as
 \begin{equation}
  \Lambda_{\rm IH;0}(u)^2
   \sim
   \kappa(u)^{2|n|N_0}\kappa(u+u_0-H(-n)I')^{2mN_0},
   \label{eq:A55}
 \end{equation}
 where $\kappa(u)$ is given by \myEqs{eq:A14}--(\ref{eq:A16})
 \cite{Baxter1978}.
 We introduce the limiting function as
 \begin{equation}
  L_{\rm IH}(u)
   =
   \lim_{N_0\rightarrow\infty}
   \frac{\Lambda_{\rm IH}(u)}
   {\kappa(u)^{2|n|N_0}\kappa(u+u_0-H(-n)I')^{2mN_0}}.
   \label{eq:A56}
 \end{equation}
 The expectation value of $\sigma_{\bf o}\sigma_{{\bf r}^+}$
 is represented as
 \begin{align}
  &\myav{\sigma_{\bf o}\sigma_{{\bf r}^+}}
  =
  \sum_p
  [{\bar{\bf A}}_0]_{0,p}
  [{\bar{\bf A}}_0]_{p,0}
  \times
  \cr
  &\qquad
  L_{{\rm IH};p}(-H(-n)I')^{2m}
  L_{{\rm IH};p}(I'-u_0)^{2|n|},
  \label{eq:A57}
 \end{align}
 where ${\bar{\bf A}}_0$ is the matrix transformed from ${\bf A}_0$
 in \myEq{eq:A19} by a similarity transformation to diagonalize
 ${\bf U}_{\rm IH}(u)$.
 Almost the same argument as in \myAppendix{appendix:A3} yields that
 $L_{\rm IH}(u)$ must be of the form
 \begin{equation}
  L_{\rm IH}(u)
   =
   \pm\prod_{l=1}^\nu
   k^\frac12\mysn(\myIm u-{\bar \phi}_l-\frac{\myIm I'}{2}), 
   \label{eq:A58}
 \end{equation}
 where
 ${\bar \phi}_l$s are
 complex numbers determined by the condition
 that the eigenvalues of the shift operator ${\bar{\bf S}}$ are
 unimodular, i.e.,
 \begin{equation}
  \left|L_{\rm IH}(-u_0)^mL_{\rm IH}(0)^n\right|=1.
   \label{eq:A59}
 \end{equation}
 From \myEqs{eq:A58} and (\ref{eq:A59}),
 we can reproduce the asymptotic behavior of the correlation function
 found in
 \myAppendix{appendix:A4}, i.e.,
 \myEq{eq:A33} or (\ref{eq:A35}) for $T>\myTc$ and
 \myEq{eq:A43} for $T<\myTc$.

 Now,
 we consider the correlation function for $T>\myTc$
 (almost the same argument holds for $T<\myTc$).
 The asymptotic correlation function is given as follows:
\begin{widetext}
 \begin{equation}
  \myav{\sigma_{\bf o}\sigma_{{\bf r}^+}}-
   \myav{\sigma_{\bf o}}\myav{\sigma_{{\bf r}^+}}
   \sim
   \int_C d{\bar \phi}{\bar \rho}({\bar \phi})
   \left[
    k^{\frac12}\mysn(-{\bar \phi}-\frac{\myIm I'}{2})
   \right]^{2m}
   \left[
    k^{\frac12}\mysn(-\myIm u_0-{\bar \phi}+\frac{\myIm I'}{2})
   \right]^{2n}.
   \label{eq:A60}
 \end{equation}
\end{widetext}
 Note that the contour $C$ is determined by the condition
 (\ref{eq:A59}), and the rotations of the lattice deform $C$.
 For instance, in the case $n=0$ ($\theta=3\pi/4$),
 $\int_C$ denotes an integral over a period interval of the length
 $2I$ on the line $\Im[{\bar \phi}]=-u_0$,
 where
 \myEq{eq:A60} reduces to \myEq{eq:A33} by the relations 
 ${\bar \phi}=\phi-\myIm u_0$ and ${\bar \rho}({\bar \phi})=\rho(\phi)$ 
 with $u_0$ replaced by $u$.
 In the case $n=m$ ($\theta=\pi/2$),
 the contour $C$ is on the line $\Im[{\bar \phi}]=-u_0/2$.
 The equivalence between \myEqs{eq:A33} and (\ref{eq:A60}) is derived
 with the help of the analyticity of the integrand.
 Using the deformation of $C$,
 we can extend the result in \myEqs{eq:A35}--(\ref{eq:A40})
 into all directions.
 
 The $\pi$ rotation of the lattice corresponds to shifting the
 integration paths by ${\rm i}I'$ in \myEq{eq:A60},
 which is connected with the fact that the
 twofold rotational
 symmetry of the
 system appears with the help of the relation
 \begin{equation}
  k^{\frac12}\mysn({\bar \phi}\pm{\rm i}I')
   =
   \left[k^{\frac12}\mysn({\bar \phi})\right]^{-1}.
   \label{eq:A61}
 \end{equation}
 In the case of isotropic interactions, where $u_0=I'/2$, 
 the $\pi/2$ rotation causes a shift of the integration paths by
 ${\rm i}I'/2$, which relates the eigenvalues of ${\bar {\bf U}}$ to
 those of $\bar{\bf S}$.
 
 \subsection{Asymptotic form for general $\boldsymbol{Q}$}
 \label{appendix:A6}

 The results by transfer matrices along various directions should be
 equivalent.
 It is pointed out that this equivalence is derived with the help of
 analytic properties of the integrand; 
 see the integrand in the right-hand side of \myEq{eq:A35},
 and also
 \myRef{Johnson1973}.
 Therefore, 
 (i) analyticity of the integrands
 is needed to ensure the equivalence between the results along various
 directions.
 Two further properties are pointed out:
 From the fact that increasing $\theta$ by $\pi$ causes $C$ to shift
 by $-\myIm I'$ along the imaginary axis, it follows that the twofold
 rotational symmetry is directly connected with the relation
 (\ref{eq:A61}).
 We find that (ii) the same relation as (\ref{eq:A61}) is satisfied by
 the limiting functions;
 see the first equation of (\ref{eq:A27}) or (\ref{eq:A28}).
 Note that
 \myEqs{eq:1.2} and (\ref{eq:1.3})
 represent elliptic curves
 (i.e., they are algebraic curves of genus~1)
 \cite{Namba1984}.
 We find that
 (iii) the asymptotic correlation function is written in terms of
 elliptic functions (or differential forms on a Riemann surface of
 genus~1).
 
 The meaning of (iii) can be explained as follows:
 Two-dimensional (2D) lattice models are related to 2D Euclidean field
 theories in their critical limit and for distances much larger than
 the lattice spacing.
 For a Euclidean field,
 the dispersion relation is written as $p_x^2+p_y^2+m^2=0$ with a
 suitable mass term $m$, and the correlation function has a periodic
 structure describing the rotational symmetry.
 For off-critical lattice models,
 two kinds of periodicity appear:
 One is connected with two-, four-, or sixfold rotational symmetry,
 and the other with the fact that eigenvalues of the transfer matrix
 are periodic functions of crystal momentum.
 This doubly periodic structure leads to the property (iii).
 
 Assuming (i)--(iii),
 we can essentially determine the leading asymptotic behavior of the
 correlation functions with the $C_{4v}$ symmetry.
 The property (iii) shows that, choosing a suitable parametrization,
 we can write the correlation function as
 \begin{equation}
  \myav{\sigma_{\bf o}\sigma_{\bf r}}-
   \myav{\sigma_{\bf o}}
   \myav{\sigma_{\bf r}}
   \sim\ {\rm Const.}
   \int^{\omega_1}_{-\omega_1}~d\Theta~
   {\cal Y}(\Theta)^j
   {\cal X}(\Theta)^i,
   \label{eq:A62}
 \end{equation}
 where 
 ${\cal Y}(\Theta)$ comes from eigenvalues of the row-to-row transfer
 matrix, and
 ${\cal X}(\Theta)$ the corresponding ones of the shift operator;
 ${\cal X}(\Theta)$ and
 ${\cal Y}(\Theta)$ are doubly periodic:
 ${\cal X}(\Theta+2\omega_1)
 ={\cal X}(\Theta+2\omega_2)
 ={\cal X}(\Theta)$
 and
 ${\cal Y}(\Theta+2\omega_1)
 ={\cal Y}(\Theta+2\omega_2)
 ={\cal Y}(\Theta)$.
 The property (ii) yields the relations 
 ${\cal Y}(\Theta+\omega_2)
 ={\cal Y}(\Theta)^{-1}$
 and 
 ${\cal X}(\Theta+\omega_2)
 ={\cal X}(\Theta)^{-1}$,
 and the property (i) indicates analyticity of
 ${\cal Y}(\Theta)$ and
 ${\cal X}(\Theta)$.
 As a result,
 ${\cal Y}(\Theta)$ and ${\cal X}(\Theta)$ must be of the forms
 \begin{equation}
  {\cal Y}(\Theta)
   =
   \prod_{l=1}^{\nu } k^{\frac12}\mysn(\Theta+\alpha_l),
   \quad%
   {\cal X}(\Theta)
   =
   \prod_{l=1}^{\nu'} k^{\frac12}\mysn(\Theta+v+\beta_l).
   \label{eq:A63}
 \end{equation}

 In the case $K=K'$ ($J=J'$) the present Ising model possesses the
 fourfold rotational symmetry.
 We can set $v=\pm\omega_2/2$, $\nu=\nu'$, and $\alpha_l=\beta_l$.
 Since the correlation function is real valued,
 we find that the modular parameter
 $\tau=\omega_2/\omega_1$ must be pure
 imaginary, which ensure the $C_{4v}$ symmetry of the system as well.
 It follows from \myEq{eq:A35} that
 the simplest case $\nu=2$ appears for $T>\myTc$.
 For parameters $\alpha_1$ and $\alpha_2$,
 we find two possibilities:
 $(\alpha_1-\alpha_2)/\omega_1$
 is purely imaginary or a real number;
 $\alpha_1-\alpha_2=\omega_2/2$
 gives \myEq{eq:A35} with $u=I'/2$. 
 The results are closely related to the $C_{4v}$ symmetry 
 except that $\alpha_1-\alpha_2=\omega_2/2$.
 We expect that \myEq{eq:A35} is applicable 
 with the relation $\alpha_1-\alpha_2=\omega_2/2$ modified suitably 
 for general $Q$ and $T>\myTc(Q)$.

 Almost the same argument holds for $T<\myTc(Q)$: 
 It follows that
 $\nu=4$,
 $\alpha_1-\alpha_2=\alpha_3-\alpha_4=\omega_2/2$, and
 $\alpha_1-\alpha_3$ is a real number.
 As mentioned above, the only difference from the
 case of $T>\myTc(Q)$ is that two elliptic curves are needed to
 represent the asymptotic correlation function
 (see \exSec{3} of \myRef{Cheng1967}
  and \exSec{XII-3} of \myRef{McCoy1973}).
  
 \section{DETAILS OF MONTE CARLO SIMULATIONS FOR INFINITE-SIZE SYSTEMS}
 \label{appendix:B}
 
 We perform large-scale MC simulations to investigate the correlation
 functions.
 In this Appendix, we shall detail our methodology.
 The Hamiltonian of the square-lattice $Q$-state Potts model is given
 by \myEq{eq:1.1}.
 We treat it in the case of $J_{{\bf r,r}'}=2J$.
 According to Fortuin and Kasteleyn (FK)
 \cite{Fortuin1972},
 the random-cluster representation of the partition function is given
 as
 \begin{align}
  Z(Q)
  &=
  \myTr{\rm e}^{-E(Q)/\mykB T}
  \cr
  &=
  \sum_{\{n\}}
  p^{\sum_{{\bf r}^\ast}n_{{\bf r}^\ast}}
  (1-p)^{\sum_{{\bf r}^\ast}(1-n_{{\bf r}^\ast})}Q^{N_{\rm c}},
  \label{eq:B1}
 \end{align}
 where $p=1-{\rm e}^{-2K}$ is the bond percolation probability.
 $n_{{\bf r}^\ast}=0, 1$ is the bond occupation defined for
 ${\bf r}^\ast\in\Lambda^*_{\rm sq}$,
 and
 $\Lambda^*_{\rm sq}$ is the medial lattice of $\Lambda_{\rm sq}$.
 We denoted the number of FK clusters as $N_{\rm c}$.
 While
 there are some variations in implementations of cluster MC simulations
 \cite{Swendsen1987,Wolff1988,Wolff1989,Evertz2001},
 we employ the so-called infinite-size method
 proposed by Evertz and von der Linden
 \cite{Evertz2001}.
 It is based on Wolff's single-cluster algorithm
 \cite{Wolff1989},
 and enables us to directly simulate infinite off-critical systems,
 which thus means that an extrapolation of data to the thermodynamic
 limit is not necessary.
 As we explained in \mySec{sec:3.a}, this advantage is crucial for our
 purpose.

 \FIGBone

 To make the explanation concrete,
 let us consider $\Lambda_{\rm sq}$ in a temperature-dependent
 bounding box of $l_{\rm B}\times l_{\rm B}$
 (see \myFig{fig:B1}).
 As an initial condition, we take random spin configurations instead
 of ``staggered spin configuration''
 \cite{Evertz2001}
 because they are neutral and unbiased for all spin states and also
 prevent a deep penetration of clusters toward the boundary (see
 below).
 We fix the seed of the cluster to the origin (the black cell) and
 perform single cluster updates in order to equilibrate a circular domain.
 Suppose that $l_{\rm T}$ is its linear dimension.
 Then,
 the required number of updates for its equilibration increases
 exponentially with $l_{\rm T}$
 because the off-critical system possesses finite correlation length $\xi$.
 Roughly speaking, 
 we performed equilibration steps to typically realize
 $l_{\rm T}\simeq20\times\xi$
 and also use the bounding box with 
 $l_{\rm B}>4\times l_{\rm T}$, where the probability that a cluster
 touches the bounding box is negligible.
 Consequently,
 we can perform measurements of the physical quantity, i.e.,
 correlation functions within the circular domain of $l_{\rm T}$
 without finite-size effects
 \cite{Evertz2001}. 

 \FIGBtwo

 With respect to the measurement of correlation functions,
 we can benefit from the so-called improved estimators.
 In the present case,
 the correlation function of the Potts model
 $c({\bf r-r}')=\myav{\sigma_{\bf r}\sigma_{{\bf r}'}}$
 can be calculated as an average over the FK clusters generated by MC,
 i.e.,
 $c({\bf r-r}')
 =
 \myav{(\sigma_{\bf r}\sigma_{{\bf r}'})_{\rm impr}}_{\rm MC}$,
 where
 \begin{equation}
  (\sigma_{\bf r}\sigma_{{\bf r}'})_{\rm impr}
   =
   \frac{1}{|C_i|}\delta({\bf r},{\bf r}'| C_i).
   \label{eq:B2}
 \end{equation}
 The set of sites (the number of sites) in a $i$th cluster were
 denoted as $C_i$ ($|C_i|$), and then
 $\delta({\bf r},{\bf r}'|C_i)=1$
 for
 ${\bf r},{\bf r}'\in{C_i}$;
 otherwise, zero.

 The correspondence between $q_{\bf r}$ and the magnetic operators is
 depicted in the right part of
 \myFig{fig:B1}.
 These magnetic operators can be characterized by the scaling
 dimensions $x(Q)$, i.e.,
 $x(2)=\frac18$, $x(3)=\frac2{15}$, $x(4)=\frac18$, and also
 $x(1)=\frac5{48}$,
 which determine the power-law behavior of $c(R)$ at $\myTc(Q)$
 \cite{Bloete1982,Deng2004}.
 The fact that \myEq{eq:B2} is positive definite is also crucial for
 calculations of vanishing correlations for large $R/\xi\gg1$.
 
 For $T>\myTc(Q)$,
 we shall provide the raw MC data of the correlation functions in two
 directions.
 In \myFig{fig:B2},
 we exhibit the semilog plots of correlation functions at various
 temperatures $t=[T-\myTc(Q)]/\myTc(Q)$.
 The pairs of blue and red lines give $\sqrt{R}c(R)$ in
 the row
 (${\bf r}=R{\bf e}_x$)
 and
 the diagonal
 [${\bf r}=R({\bf e}_x+{\bf e}_y)/\sqrt{2}$]
 directions.
 Then, 
 one finds that
 their slopes become steeper, and
 the discrepancy of the pair of lines becomes larger with the increase
 of the reduced temperature $t$.
 For exactly solved cases,
 it was revealed that the correlation length is isotropic near
 critical point, but becomes anisotropic at a distance from it
 due to the lattice effects.
 With this in mind, 
 if we suppose the Ornstein-Zernike form of the correlation function as
 $c(R)\propto{\rm e}^{-R/\xi}/\sqrt{R}$,
 then our MC data indicate that $\xi$ in the row direction
 is longer than that in the diagonal direction.
 Simultaneously,
 one can notice that the directional dependence of $\xi$ is quite
 weak,
 so the extremely accurate data are required to investigate the
 $Q$ dependence of the ACLs.
 
 \section{%
 FITTING CALCULATION OF THE FORM FOR MONTE CARLO DATA}
 \label{appendix:C}
 
 In this Appendix,
 we detail a fitting procedure of our form (\ref{eq:3.2}) to the
 correlation function data provided by the MC simulation calculations.
 As explained in \myAppendix{appendix:B},
 the infinite-size MC method and the improved estimator for the
 correlation functions has been employed.
 In typical cases,
 we performed 1000 independent runs of MC simulations and generated
 the $10^{11}$ Fortuin-Kasteleyn clusters at each run.
 Then,
 for square-lattice sites
 $i{\bf e}_x+j{\bf e}_y\in\Lambda_{\rm sq}$
 within the equilibrated circular domain
 $R<l_{\rm T}$
 the correlation function data
 $\{c(i,j)\}$
 were obtained, and their statistical errors
 $\{d(i,j)\}$
 were estimated from standard deviations of the averages of the
 independent runs.
 
 As mentioned in \mySec{sec:3.a},
 there exist two sources of errors:
 the systematic errors stemming from higher bands of eigenvalues which
 are not taken into account in \myEq{eq:3.2} and the statistical
 errors associated with MC samplings.
 The former
 (respectively, latter)
 becomes larger inward
 (respectively, outward).
 We analyze $\{c(i,j)\}$ and $\{d(i,j)\}$ in annular regions
 $\myann(c_{\rm max},c_{\rm min})$ following the procedure explained
 below.
 
 We shall take the $Q=3$ and $t=0.15$ case as an example.
 In \mySec{sec:3.c},
 we order-estimated the systematic error as
 $\Delta(3)\times{\rm e}^{-2R/\bar\xi}$
 with
 ${\bar \xi}\simeq 2.7$
 and 
 $\Delta(3)=1.0-({A}/{\pi})(1-k^2)^{\frac14}\times2I\simeq O(10^{-2})$.
 Therefore,
 to calculate the ACL within a five-digit accuracy,
 we need to employ an annular region with mean radius $\simeq20$ or
 longer.
 Because the statistical errors are larger in outer regions, 
 we choose
 $\myann_0=\myann(1\times 10^{-4},3\times10^{-5})$
 with mean radius
 $R_0\simeq 21.3$
 as an optimized region.

 \TABLECone

 Then,
 we define $\chi^2$ as a function of $A,~k,~b$ by
 \begin{equation}
  \chi^2(A,k,b)
   =
   \sum_{(i,j)\in\myann_0}
   \left[
    \frac{{\cal F}_{\rm sq}(i,j;A,k,b)-c(i,j)}{d(i,j)}
   \right]^2,
   \label{eq:C1}
 \end{equation}
 We extract values $\bar A$, $\bar k$, and $\bar b$ 
 of the fitting parameters by minimizing $\chi^2(A,k,b)$.
 The first line of the third column in \myTable{table:1} gives their
 estimates and errors given by the parenthesized digits, 
 which were put based on differences between two results to two groups
 of independent runs (e.g., we divided 1000 independent runs into two
 groups and performed fitting calculation for each).
 
 We have expected the extracted values to be obtained under a
 well-controlled condition of systematic errors by carefully choosing
 the fitting region.
 To show concrete evidence to this statement, 
 we perform fittings of data in different annular regions
 $\myann_\alpha$ and check the $\myann_\alpha$ dependence of an
 estimate as well as a reduced $\chi^2$ values, i.e.,
 \begin{equation}
  \bar\chi^2_{|\myann_\alpha|}
   =
   \frac{\chi^2(\bar A,\bar k,\bar b)}{|\myann_\alpha|}.
   \label{eq:C2}
 \end{equation}
 The third column of \myTable{table:C1} compares the estimates in
 one inner region
 ($\myann_{-1}$),
 the optimized region
 ($\myann_0$), and
 two outer regions ($\myann_1$ and $\myann_2$).
 In general,
 $\bar\chi^2_{|\myann_\alpha|}$ measures the goodness of fit, which in the
 present case gives an applicability condition of \myEq{eq:3.2} to MC
 data in $\myann_\alpha$.
 First,
 one sees that
 $\bar\chi^2_{|\myann_{-1}|}$ is much larger than the others,
 and that
 $\xi_{\rm diag}$ estimated in $\myann_{-1}$ deviates largely from
 those in other regions.
 Meanwhile,
 the error in $\xi_{\rm diag}$ shown by parenthesized digits becomes
 smaller for $\myann_{-1}$.
 These show that \myEq{eq:3.2} cannot fit the data in $\myann_{-1}$
 due to the systematic errors.
 Second,
 one also finds that
 $\bar\chi^2_{|\myann_1|}$ and $\bar\chi^2_{|\myann_2|}$ are
 comparable to $\bar\chi^2_{|\myann_0|}$,
 and that
 the estimates of $\xi_{\rm diag}$ are almost independent of the
 choice of the outer regions.
 Therefore,
 we conclude that
 $\myann_0$, $\myann_1$, and $\myann_2$ are in an asymptotic region in
 which \myEq{eq:3.2} can be used for the fitting under the controlled
 condition of systematic errors, but the statistical errors become
 larger in outer region.

 The first,
 the second,
 and
 the fourth
 columns of \myTable{table:C1} give
 the results obtained via the same analysis for
 the $Q=1,\ t=0.50$, 
 the $Q=2,\ t=0.24$, 
 and
 the $Q=4,\ t=0.10$
 cases, respectively.
 While the overall feature of $Q=1$, 4 is same as that of $Q=3$,
 the fitting condition for $Q=2$ is clearly different from them,
 namely,
 both $\bar\chi^2_{|\myann_{\alpha}|}$ and $\xi_{\rm diag}$ are almost
 independent of $\myann_{\alpha}$.
 This difference can be attributed to the presence/absence of the
 second band eigenvalue contributions to the correlation function:
 As explained in \myAppendix{appendix:A},
 they are absent in the Ising case so that \myEq{eq:3.2} can fit the
 data in inner annular regions like $\myann_{-1}$.

 \myTable{table:1} exhibits the fitting data in optimized regions
 $\myann_0$ for which the convergence check of estimates as
 demonstrated in \myTable{table:C1} has been performed at all
 temperatures.
 In principle, 
 we can employ a wider annular region including, e.g., $\myann_0$,
 $\myann_1$, and $\myann_2$,
 but, in reality
 the infinite-size algorithm MC simulations cannot provide
 reliable averages and meaningful errors for $R\gg\xi_{\rm diag}$
 within a moderate computational effort
 \cite{Evertz2001}.
 Therefore,
 the optimization of the fitting region is necessary for the
 purpose of
 the accurate estimations of the ACLs.


 \end{document}